\documentclass{article}
\usepackage{graphicx}
\graphicspath{Fig/}

\usepackage[utf8]{inputenc}
\usepackage[margin=1in]{geometry}
\usepackage{color}
\usepackage{enumitem}
\usepackage{authblk}

\usepackage{hyperref}
\usepackage{hypcap}

\usepackage[ruled,vlined]{algorithm2e}
\usepackage{algorithmic}

\usepackage{float}
\usepackage{caption}
\usepackage{subcaption}
\usepackage{setspace}
\usepackage{tcolorbox}

\usepackage{amsmath,amsfonts,amssymb,mathtools, amsthm}

\usepackage[english]{babel}

\newtheorem{remark}{Remark}
\numberwithin{equation}{section}

\title{
Runaway electron control by self-excited waves
}

\author[1]{Kun Huang\thanks{Email: \texttt{kunhuang@vt.edu}}}
\author[2]{Boris Breizman\thanks{Email: \texttt{breizman@mail.utexas.edu}}}
\affil[1]{Department of Mathematics, Virginia Polytechnic Institute and State University, VA 24061, USA}
\affil[2]{Institute for Fusion Studies, The University of Texas at Austin, TX 78712, USA}

\date{}

\begin{document}
\maketitle

\begin{abstract}

Runaway-electron avalanches in tokamak plasmas can be limited
by kinetic instabilities driven by the non-Maxwellian runaway distribution. We formulate a reduced model for the quasi-steady state in which the total plasma current and bulk electron temperature are prescribed, while the inductive electric field is determined self-consistently from the partition between Ohmic bulk current and runaway-electron current. Because the wave growth time is short compared with the current-decay time, we consider a marginal-stability regime, in which whistler-wave drive by the runaway electrons balances collisional damping.

The resulting states separate
into three regimes: a subcritical Ohmic regime without an avalanche, an avalanche regime in which runaway growth relaxes the inductive field to the avalanche threshold, and an instability-regulated regime in which self-excited whistler waves enhance momentum-space diffusion and limit the runaway current.

In the instability-regulated regime, the whistler wave spectrum forms a narrow ridge, and low-energy runaway electrons carry most of the runaway current.  
\end{abstract}

\section{Introduction}
When the driving electric field is large enough to overcome the drag force caused by Coulomb collision, a group of electrons will accelerate to relativistic energies, which is known as an electron runaway effect. 

The drag force due to synchrotron radiation can limit the electron energy gain in magnetically confined plasmas while elastic scattering spreads electrons in pitch angle. An electron accelerates to relativistic energy, scatters due to Coulomb collision with ions, increases its pitch angle, and slows down because of  synchrotron losses.  These processes create a loop in momentum space. Without an extra source, the runaway electron (RE) population will gradually decay as some electrons leak into the bulk plasma area diffusively. The decay rate of the runaway population depends on the driving field because a stronger field shifts the population to  higher energies.

Production of new runaway electrons via knock-on collisions of the runaway population with the bulk can overcome the diffusive leak. In that case, the runaway population will grow exponentially as an avalanche. The rate of knock-on collisions is lower than the small-angle collision frequency. It can, nevertheless, compete with the diffusive losses when the driving field is sufficiently strong. The minimal driving field required to overcome the losses determines the avalanche threshold. 

The shape of the RE distribution does not change significantly during the slow-growing avalanche. In other words,  the RE population and the runaway current simply grow in amplitude in a quasi-steady way. 

A non-Maxwellian distribution of the runaway electrons is prone to high-frequency kinetic instabilities when the runaway current is large enough to overcome the collisional damping of the excited waves. These damping rates do not depend on the RE parameters, whereas the instability drive increases linearly with the runaway current. Instabilities can develop before the growing runaway electron current replaces the total plasma current, in which case the feedback from the excited waves should modify the runaway population and may even stop the avalanche when the runaway current reaches the instability threshold. Alternatively, in the case of subthreshold total current, the avalanche will stop without exciting any instability. That would be due to the decrease of the driving inductive electric field when the runaways replace a significant part of the total current.

The characteristic growth times for the instabilities of interest are
much shorter than the lifetime of the runaway current, which suggests that an initially unstable system will relax to a marginally stable state, in which a quasi-steady spectrum of waves determines the shape of the runaway electron population and limits the runaway current. In such a state, the knock-on collisions must balance the diffusive losses of the  runaway electrons. The lifetime of this state would be determined by the dissipation of magnetic energy associated with the slow decay of the total plasma current.   

This paper analyzes the quasi-steady runaway-electron distribution and its dependence on the total current and the bulk electron temperature. Several earlier studies provide complementary perspectives on this problem. Liu et. al. \cite{liu2018role} considered the kinetic evolution at a prescribed electric field, whereas in our paper the steady state is determined at a prescribed total current, with the inductive electric field obtained self-consistently by the partition between bulk and runaway current. One important observation we have from the numerical experiments is that even for large total current, the inductive field at the final quasi-steady state does not exceed the Connor-Hastie field very much. The paper of Breizman and Kiramov \cite{breizman2023marginal} gave the distribution of wave-controlled RE assuming high energy and small pitch-angle. Their calculation showed that runaway electrons tend to accumulate at the low energy region of the momentum space. However, the predicted distribution in that paper blows up at zero, because the high-energy and small-angle assumptions are no longer valid there. Our work covers the whole momentum space by solving a more sophisticated model numerically. It is worth noting that the quasilinear theory used in the aforementioned works, including ours, assumes that wave modes are dense enough to be treated as a continuous spectrum. In some scenarios this is not true, and the model has to consider discrete modes of waves, which has been discussed in \cite{breizman2023nonlinear}.

The paper will be organized as follows. Section \ref{sec:model} introduces the kinetic model for the coupled runaway-electron and wave dynamics. Section \ref{sec:margin} reduces the problem to a marginally stable state with a self-consistent inductive field and a one-dimensional approximation to the wave spectrum. Section \ref{sec:experiment} then uses numerical experiments to identify the resulting current regimes, temperature dependence, and wave-spectrum structure, followed by conclusions in Section \ref{sec:conclusion}.

\section{Kinetic model} \label{sec:model}
We use a reduced, gyro-averaged kinetic model on a fixed flux surface, following the structure of the runaway-electron kinetic equation reviewed in \cite{breizman2019physics}. Finite-orbit-width effects, radial transport, bremsstrahlung, and pair production are not included. The retained ingredients are: the inductive electric field, Coulomb drag, pitch-angle scattering, synchrotron radiation reaction, knock-on production, and quasilinear diffusion caused by self-excited waves. The model evolves the runaway-electron distribution $f(\mathbf{p},t)$ and the wave spectral energy density $W(\mathbf{k},t)$:
\begin{equation}
    \begin{split}
        \partial_{t}f=&\mathbb{E}f+\mathbb{C}f+\mathbb{Z}f+\mathbb{R}f+\mathbb{S}f +\mathbb{D}[W]f,\\
        \partial_{t}W =& \left( 2\Gamma_{b}[\nabla_{p} f] - 2\Gamma_{\nu}\right)W + \Sigma.
    \end{split}
\end{equation}
The operators $\mathbb{E}$, $\mathbb{C}$, $\mathbb{Z}$, $\mathbb{R}$, $\mathbb{S}$, and $\mathbb{D}$ denote, respectively, electric-field acceleration, collisional drag on electrons, pitch-angle scattering on ions, synchrotron radiation reaction, knock-on source, and quasilinear diffusion. The equation for waves contains the kinetic drive from RE $\Gamma_b$, the collisional damping rate $\Gamma_\nu$, and the thermal-noise source $\Sigma$. 

Let $\mathbf{P} = (P_{\parallel}, P_{\perp})$ be the momentum normalized by $mc$, and let $P = \sqrt{P_{\parallel}^{2} + P_{\perp}^{2}}$. Thus $\mathbf{P}=\mathbf{p}/mc$, $\gamma=\sqrt{1+P^2}$, $v_\parallel=p_\parallel/\gamma m$, $v_\perp=p_\perp/\gamma m$, $\cos\theta=P_\parallel/P$, and $\sin\theta=P_\perp/P$. We take $e>0$ to be the elementary charge, $m$ the electron mass, $c$ the speed of light, $B$ the background magnetic-field strength, $n_e$ the electron density, $n_{ion}$ the ion density, $Z$ the ion charge number, and $\ln\Lambda$ the Coulomb logarithm. The partial contributions to the RE kinetic equation are defined as follows:
\begin{itemize}
    \item Drag force due to Coulomb collision with bulk electrons:
    \begin{equation}\label{eq:edrag}
        \begin{split}
            \mathbb{C}f=&\frac{\partial}{\partial p_{\parallel}}\left(eE_{CH}\frac{\gamma^{2}m^{2}c^{2}}{p^{2}}\frac{p_{\parallel}}{p}f\right)+\frac{1}{p_{\perp}}\frac{\partial}{\partial p_{\perp}}\left(p_{\perp}eE_{CH}\frac{\gamma^{2}m^{2}c^{2}}{p^{2}}\frac{p_{\perp}}{p}f\right)\\
            =&\frac{1}{\tau_{CH}}\frac{\partial}{\partial P_{\parallel}}\left(\frac{\gamma^{2}}{P^{2}}\frac{P_{\parallel}}{P}f\right)+\frac{1}{\tau_{CH}}\frac{1}{P_{\perp}}\frac{\partial}{\partial P_{\perp}}\left(P_{\perp}\frac{\gamma^{2}}{P^{2}}\frac{P_{\perp}}{P}f\right),
        \end{split}
    \end{equation}
    where the Connor-Hastie threshold $E_{CH} = \frac{4\pi e^{3}n_{e}\ln\Lambda}{mc^{2}}$ in cgs units \cite{breizman2019physics} (in what follows, every constant is given in cgs units by default). And the associated time scale $\tau_{CH} = \frac{mc}{eE_{CH}}=\frac{m^{2}c^{3}}{4\pi e^{4}n_{e}\ln\Lambda}$.
    
    \item Elastic pitch-angle scattering on ions: 
    \begin{equation}
            \mathbb{Z}f=\frac{1}{\tau_{Z}}\frac{1}{P^{2}\sin\theta}\partial_{\theta}\left(\frac{\gamma}{P}\sin\theta\partial_{\theta}f\right),
    \end{equation}
    where $\frac{1}{\tau_{Z}} = \frac{2\pi e^{4}}{m^{2}c^{3}}n_{ion}Z^{2}\ln\Lambda$ \cite{breizman2019physics} and
    \begin{equation*}
        \frac{\tau_{Z}}{\tau_{CH}} = \frac{2n_{e}}{Z^{2}n_{ion}}.
    \end{equation*}

    \item Electric field drive:
    \begin{equation}
        \mathbb{E}f = -\frac{1}{\tau_{CH}}\frac{\partial}{\partial P_{\parallel}}\left(\frac{E(t)}{E_{CH}}f\right).
    \end{equation}
    where $E(t)$ will be given in Equation \ref{eq:inductive_E}.

    \item Synchrotron radiation:
    \begin{equation}
        \mathbb{R}f=\frac{1}{\tau_{R}}\left(\frac{\partial}{\partial P_{\parallel}}\left(\frac{P_{\parallel}P_{\perp}^{2}}{\sqrt{1+P^{2}}}f\right)+\frac{1}{P_{\perp}}\frac{\partial}{\partial P_{\perp}}\left(P_{\perp}\frac{P_{\perp}\left(1+P_{\perp}^{2}\right)}{\sqrt{1+P^{2}}}f\right)\right),
    \end{equation}
    where $\tau_{R}=\frac{3m^{3}c^{5}}{2e^{4}B^{2}}$ \cite{breizman2019physics} and $\frac{\tau_{R}}{\tau_{CH}} = \frac{6\pi n_{e}mc^{2}}{B^{2}}\ln\Lambda$. Note that $\omega_{ce}=\frac{eB}{mc}$ and $\omega_{pe}^{2}=\frac{4\pi n_{e}e^{2}}{m}$, therefore
    \begin{equation*}
        \frac{\tau_{R}}{\tau_{CH}}=\frac{3}{2}\frac{\omega_{pe}^{2}}{\omega_{ce}^{2}}\ln\Lambda.
    \end{equation*}

    \item Knock-on collision:
    \begin{equation}
        \mathbb{S}f = \int f(\mathbf{P}_{0})\mathcal{K}(\mathbf{P},\mathbf{P}_{0})1_{\left\{ \gamma>\gamma_{m},1+\gamma_{0}-\gamma>\gamma_{m}\right\} }1_{\left\{ \gamma_{0}>2\gamma-1\right\} }d^{3}\mathbf{P}_{0},
    \end{equation}
    where the collisional kernel \cite{breizman2019physics} takes the following form:
    \begin{equation*}
        \mathcal{K}(\mathbf{P},\mathbf{P}_{0})=\frac{n_{e}c}{2\pi}\frac{P_{0}}{P}\frac{1}{\gamma_{0}}\frac{1}{\gamma}\left(\frac{d\sigma}{d\gamma}\right)_{\gamma\gamma_{0}}\delta\left(\mathbf{n}\cdot\mathbf{n}_{0}-\sqrt{\frac{\gamma-1}{\gamma+1}}\sqrt{\frac{\gamma_{0}+1}{\gamma_{0}-1}}\right),
    \end{equation*}
    and
    \begin{equation*}
        \left(\frac{d\sigma}{d\gamma}\right)_{\gamma\gamma_{0}}=\frac{2\pi r_{e}^{2}}{\gamma_{0}^{2}-1}\left[\gamma_{0}^{2}\left(\frac{1}{\gamma-1}\right)^{2}+\gamma_{0}^{2}\left(\frac{1}{\gamma_{0}-\gamma}\right)^{2}+1-\frac{2\gamma_{0}-1}{\gamma_{0}-1}\left(\frac{1}{\gamma-1}+\frac{1}{\gamma_{0}-\gamma}\right)\right].
    \end{equation*}

    In this expression $\mathbf{P}_0$ denotes the incoming runaway electron momentum, $\gamma_{0} = \sqrt{1+P_{0}^{2}}$ is the corresponding Lorentz factor, $\mathbf{n}=\mathbf{P}/P$ and $\mathbf{n}_0=\mathbf{P}_0/P_0$ are the corresponding direction vectors, $\delta$ is the Dirac delta function, and $1_{\{\cdot\}}$ is an indicator function. The cut-off parameter $\gamma_m$ defines the lower energy boundary of the runaway region, and $(d\sigma/d\gamma)_{\gamma\gamma_0}$ is the Moller differential cross section for producing an electron with Lorentz factor $\gamma$ from an incident electron with Lorentz factor $\gamma_0$. The angular delta function enforces the two-body scattering kinematics.

    Let $\frac{1}{\tau_{S}} = \frac{n_{e}c}{2\pi}\cdot2\pi r_{e}^{2} = n_{e} c r_{e}^2$. The classical electron radius $r_{e} = \frac{e^{2}}{mc^{2}}$, hence $\frac{\tau_{S}}{\tau_{CH}} = 4\pi\ln\Lambda$.
\end{itemize}

The quasilinear diffusion operator $\mathbb{D}f$ describes resonant interaction between runaway electrons and the excited waves. We use the divergence form \cite{breizman2019physics}:
\begin{equation}\label{eq:operator_D}
    \mathbb{D}f=\nabla_{p}\cdot\left(\mathcal{D}[W]\cdot\nabla_{p}f\right),
\end{equation}

where the diffusion tensor $\mathcal{D}$ is linear in $W(\mathbf{k},t)$:

\begin{equation}\label{eq:qlt_diffusion}
    \begin{split}
        \mathcal{D}[W] &= \sum_{l} \int d^{3}\mathbf{k}\left(\beta\otimes\beta\right) W(\mathbf{k}, t) U_{l}(\mathbf{p},\mathbf{k})\delta(\omega-k_{\parallel}v_{\parallel}-\frac{l\omega_{c}}{\gamma}),\\
        \beta &\coloneqq \frac{k_{\parallel}v_{\parallel}}{\omega}\frac{p}{p_{\parallel}}\mathbf{e}_{\parallel} + (1-\frac{k_{\parallel}v_{\parallel}}{\omega})\frac{p}{p_{\perp}}\mathbf{e}_{\perp}.
    \end{split}
\end{equation}
Here $\mathbf{k}$ is the wave vector, $\omega=\omega(\mathbf{k})$ is the wave dispersion relation, $l$ is the cyclotron-harmonic index, and $\omega_c$ is the electron cyclotron frequency. The factor $U_l(\mathbf{p},\mathbf{k})$ is the harmonic-resolved wave-particle coupling kernel, as given in Eq. (133) of \cite{breizman2019physics}. The vector $\beta$ is the direction in momentum space along which the resonant wave-particle interaction produces diffusion.

The same resonant interaction drives or damps the waves according to
\begin{equation}
    \partial_{t}W = \left( 2\Gamma_{b}[\nabla_{p} f] - 2\Gamma_{\nu}\right)W + \Sigma,
\end{equation}
where the beam-driven growth rate $\Gamma_{b}$ is linear in the momentum-space gradient of the runaway distribution:
\begin{equation}\label{eq:magreaction}
    \Gamma_{b}(\mathbf{k},t)= \frac{1}{2} \sum_{l}\int d^{3} \mathbf{p} \left(\beta\cdot\nabla_{p}f\right)\left(\beta \cdot \nabla_{p}\gamma mc\right)U_{l}(\mathbf{p},\mathbf{k})\delta(\omega-k_{\parallel}v_{\parallel}-l\omega_{c}/\gamma).
\end{equation}

The collisional damping rate is solely determined by plasma parameters:
\begin{equation*}
    \Gamma_{\nu}
=
\nu_{ei}
\frac{
E_\alpha^{*}E_\beta\,\omega\,\dfrac{\partial}{\partial\omega}
\left[\omega\left(\varepsilon^{H}_{\alpha\beta}-\delta_{\alpha\beta}\right)\right]
}{
E_\alpha^{*}E_\beta \dfrac{\partial}{\partial\omega}
\left(\omega^{2}\varepsilon^{H}_{\alpha\beta}\right)
},
\end{equation*}
with
\begin{equation*}
    \nu_{ei}
=
\frac{4\sqrt{2\pi}\,e^{4}\ln\Lambda}
{3m^{1/2}T_{e}^{3/2}}
Z^{2}n_{ion} .
\end{equation*}
In the damping formula, repeated Cartesian polarization indices $\alpha,\beta$ are summed, $E_\alpha$ is the wave electric-field polarization vector, the star denotes complex conjugation, $\varepsilon^H_{\alpha\beta}$ is the Hermitian part of the cold-plasma dielectric tensor, and $\delta_{\alpha\beta}$ is the Kronecker delta. The electron-ion collision frequency is $\nu_{ei}$, $T_e$ is the bulk electron temperature, and $\tau_\nu=1/\nu_{ei}$ is the corresponding collision time.

\begin{equation*}
    \frac{\tau_{\nu}}{\tau_{CH}} = \frac{1}{\nu_{ei}\tau_{CH}} = \frac{1}{\left(\frac{mc^{2}}{T_{e}}\right)^{3/2}\frac{2 Z^{2}n_{ion}}{3\sqrt{2\pi} n_{e}}} = \frac{3\sqrt{2\pi}}{2}\left(\frac{T_{e}}{mc^{2}}\right)^{3/2}\frac{ n_{e}}{ Z^{2}n_{ion}}
\end{equation*}

The source $\Sigma(\mathbf{k})$ represents the thermal fluctuation level that seeds the wave spectrum. We assume that $\Sigma/\Gamma_{\nu}$ is smooth in $\mathbf{k}$ and much smaller than the level of turbulence excited by the runaway beam. The marginally stable states studied below are insensitive to the detailed form of this seed spectrum.

The cold-plasma dielectric tensor $\varepsilon_{\alpha \beta}$, wave polarization vector $E_{\alpha}$, and explicit expression for $U_l$ are collected in Appendix \ref{sec:appendix_kinetic_instability}.

\section{Marginally stable state} \label{sec:margin}

\subsection{The self-consistent inductive field}

During disruption, the total current in a tokamak tends to decrease as the resistivity $\eta$ increases. Due to large inductance, a strong inductive electric field is generated, trying to restore the current. As a result, the total current decreases so slowly that it can be treated as a time-independent constant on the kinetic relaxation time scale considered here.

The total current consists of the bulk-electron and runaway-electron contributions. In current form, $I=I_{RE}+I_{bulk}$. In the rest of the paper we use the corresponding current densities,
\begin{equation*}
    j_{tot}=j_{RE}(t)+j_{bulk}(t).
\end{equation*}
Assume the bulk current to be Ohmic, i.e.
\begin{equation*}
    j_{bulk}(t)=\frac{E(t)}{\eta(t)},
\end{equation*}
it follows that 
\begin{equation}\label{eq:inductive_E}
    E(t)=\eta(t)\left[j_{tot}-j_{RE}(t)\right].
\end{equation}

\subsection{Steady RE distribution depending on total current}
With the inductive field $E[f]$ given in Equation \ref{eq:inductive_E}, we now have all the ingredients of the kinetic equations:
\begin{equation*}
    \begin{split}
        \partial_{t}f=&\mathbb{E}[f]f+\mathbb{C}f+\mathbb{Z}f+\mathbb{R}f+\mathbb{S}f +\mathbb{D}[W]f,\\
        \partial_{t}W =& \left(2 \Gamma_{b}[\nabla_{p} f] - 2\Gamma_{\nu}\right)W + \Sigma.
    \end{split}
\end{equation*}

As the runaway population grows, its non-Maxwellian momentum-space gradient can drive kinetic instabilities. The resulting waves enhance momentum-space diffusion and increase the loss of runaways back to the bulk. We therefore seek the quasi-steady distribution $f_\infty$ and runaway current density $j_{RE}$ selected by a prescribed total current density $j_{tot}$.

Since we are only interested in the final state, it is unnecessary to run a simulation with self-consistent electric field given in Equation \ref{eq:inductive_E}. Instead, assuming that the resistivity $\eta(t)$ is almost time-independent, we propose the following experiment procedure to find the relation $f_{\infty}(j_{tot})$.

\begin{enumerate}
    \item For a series of supercritical constant fields $E$, find the steady states $f_{*}(E)$ and $W_{*}(E)$ such that
    \begin{equation*}
        \begin{split}
            \mathbb{E}f_{*}+\mathbb{C}f_{*}+\mathbb{Z}f_{*}+\mathbb{R}f_{*}+\mathbb{D}f_{*}+\mathbb{S}f_{*} &= 0\\
            \left(2 \Gamma_{b}[\nabla_{p} f_{*}] - 2 \Gamma_{\nu}\right)W_{*} + \Sigma &= 0
        \end{split}
    \end{equation*}
    \item Perform curve fitting to obtain the relation $j_{RE}=X(E)$, which should be monotonically increasing. It follows that
    \begin{equation*}
        j_{tot} = \frac{E}{\eta} + X(E),
    \end{equation*}
    which should also be monotonically increasing. Hence the inverse function $Y=F^{-1}$, where $F(E)=E/\eta+X(E)$, exists and
    \begin{equation*}
        E = Y(j_{tot}).
    \end{equation*}
\end{enumerate}

It can be easily proved that for any given total current density $j_{tot}$, let $E=Y(j_{tot})$, then $f_{*} = f_{*}(E)$ is exactly the steady state of the kinetic equation with self-consistent field:
\begin{equation*}
    \begin{split}
        \mathbb{E}[f_{*}]f_{*}+\mathbb{C}f_{*}+\mathbb{Z}f_{*}+\mathbb{R}f_{*}+\mathbb{D}f_{*}+\mathbb{S}f_{*} &= 0,\\
            \left( 2\Gamma_{b}[\nabla_{p} f_{*}] - 2\Gamma_{\nu}\right)W_{*} + \Sigma &= 0,
    \end{split}
\end{equation*}
thus we have found the relation between $f_{\infty}$ and $j_{tot}$.

The Spitzer resistivity is inversely proportional to $T_{e}^{3/2}$, with $T_{e}$ being the bulk electron temperature:
\begin{equation*}
    \eta = \frac{4\sqrt{2\pi}}{3}\frac{Z e^{2}\sqrt{m} \ln{\Lambda}}{T_{e}^{3/2}}.
\end{equation*}
Recall that the Connor-Hastie field is:
\begin{equation*}
    E_{CH} = \frac{4\pi e^{3}n \ln{\Lambda}}{mc^{2}}, 
\end{equation*}
hence the typical bulk current density is in the order of
\begin{equation*}
    \frac{1}{\eta}E_{CH} = \frac{3\sqrt{2\pi}}{2}\frac{1}{Z}\left(\frac{T}{mc^{2}}\right)^{3/2}nec.
\end{equation*}

When the bulk electron number density $n = 1.0\times 10^{20} m^{-3}$, bulk temperature $T_{e} = 18 eV$ and the effective charge of impurities $Z = 6$, we have 
\begin{equation*}
    \frac{1}{\eta}E_{CH} = 650 A/m^{2}.
\end{equation*}

For comparison, the typical total current density in a tokamak is around $1.0 \times 10^{6} A/m^{2}$, which means at low bulk temperature, in the end almost all of the current comes from runaway electrons.

\subsection{Quasilinear diffusion operator at marginal stability}
The characteristic growth times for the instabilities of interest are much shorter than the lifetime of the runaway current, therefore it is reasonable to assume that the wave spectral energy density reaches equilibrium in no time. For simplicity, assuming that the thermal noise term $\Sigma = 2\sigma \Gamma_{\nu}$ with $\sigma$ being a small constant, and denote the ratio between excitation $\Gamma_{b}$ and damping $\Gamma_{d}$ as $R$, we have
\begin{equation}\label{eq:W_marginal}
    W[f] = \frac{\Sigma}{2\Gamma_{\nu}-2\Gamma_{b}[\nabla_{p}f]} = \sigma\frac{1}{1-\Gamma_{b}[\nabla_{p}f]/\Gamma_{\nu}} = \frac{\sigma}{1-R[f]}.
\end{equation}
Substitute it in \eqref{eq:operator_D}, the definition of quasilinear diffusion operator $\mathbb{D}$, we obtain the self-consistent operator $D[f]$. Note that for a given parameter $\sigma$, there is a one-to-one relation between $R$ and $W$, hence we will call the function $R(\mathbf{k})$ as the "marginality spectrum" in the rest of this paper.

\subsubsection{The one-dimensional ridge assumption on the wave spectrum}
The key assumption of this work is as follows: at marginal stability, the excitation $\Gamma_{b}$ and damping $\Gamma_{\nu}$ reach a balance along a one-dimensional curve in the two-dimensional $(k_{\parallel}, k_{\perp})$ space. This assumption will be justified a posteriori through numerical experiments. Based on this assumption, the wave spectrum can be approximated with the following form:
\begin{equation*}
    W(k_{\parallel}, k_{\perp}) = A(k_{\parallel})\delta(k_{\perp} - B(k_{\parallel})).
\end{equation*}
Substitute it into \eqref{eq:qlt_diffusion}, we obtain
\begin{equation*}
    \begin{split}
        \mathcal{D} =& \sum_{l} \iint 2\pi k_{\perp} dk_{\perp} dk_{\parallel}\left(\beta\otimes\beta\right) A(k_{\parallel})\delta(k_{\perp} - B(k_{\parallel})) U_{l}(\mathbf{p},\mathbf{k})\delta(\omega-k_{\parallel}v_{\parallel}-\frac{l\omega_{c}}{\gamma})\\
        =& \sum_{l} \int 2\pi A(k_{\parallel}) B(k_\parallel) \left(\left(\beta\otimes\beta\right)U_{l}\right)(\mathbf{p}, k_\parallel, B(k_\parallel))\delta(\omega(k_\parallel, B(k_\parallel))-k_{\parallel}v_{\parallel}-\frac{l\omega_{c}}{\gamma}) dk_{\parallel}
    \end{split}
\end{equation*}

\begin{remark}
    Indeed, the one dimensional spectrum is the only possible configuration to reach a steady state. If $R[f] \approx 1$ inside a 2D subset of the $(k_{\parallel},k_{\perp})$ space, the integral blows up and there will be infinite diffusion in momentum space; if $R[f] \approx 1$ only at discrete 0D points, the integral will be equal to zero, which means no enhanced diffusion at all.
\end{remark}

\bigskip
It remains to find the mapping from $W$ to $A$ and $B$. On each $k_{\parallel}$, we choose $B = \arg \max W(k_{\perp})$ as the location of the delta spectrum. To match the total energy, let
\begin{equation*}
    \int W(k_{\perp}) 2\pi k_{\perp} d k_{\perp} = \int A \delta(k_{\perp} - B) 2\pi k_{\perp} d k_{\perp},
\end{equation*}
hence we need
\begin{equation*}
    A = \int \frac{W(k_{\perp})}{B} k_{\perp} d k_{\perp}.
\end{equation*}

\subsubsection{Regularized wave response}
For a prescribed distribution, the formal wave-energy balance in the linear wave equation gives
\begin{equation}\label{eq:orginalW}
    W(\mathbf{k})
    =
    \frac{\Sigma(\mathbf{k})}{2\Gamma_\nu(\mathbf{k})-2\Gamma_b(\mathbf{k})}
    =
    \frac{\sigma}{1-R(\mathbf{k})},
    \qquad
    R(\mathbf{k})=\frac{\Gamma_b(\mathbf{k})}{\Gamma_\nu(\mathbf{k})},
    \qquad
    \sigma=\frac{\Sigma(\mathbf{k})}{2\Gamma_\nu(\mathbf{k})}.
\end{equation}
    
The singular factor $(1-R)^{-1}$ identifies where the wave spectrum is selected: waves become important only near marginality, $R\simeq1$. However, a solver for steady states based on fixed-point iteration using the above formula suffers from very slow convergence. Therefore, we propose a novel technique to accelerate the numerical solver by replacing \eqref{eq:orginalW} with the sigmoid function:
\begin{equation*}
    W(\mathbf{k})
    =
    \frac{W_{\max}}{1+\exp[-\alpha(R(\mathbf{k})-1)]}.
\end{equation*}
This function is an implementation-level approximation to the singular response $\sigma/(1-R)$: it is small well below marginality, turns on in a narrow layer around $R=1$, and saturates at a finite cap once the drive exceeds the damping. The parameters $W_{\max}$ and $\alpha$ therefore regularize the amplitude and transition width, while the location of the active spectrum is still determined by the marginal-stability condition.

\bigskip
In what follows, we will explain why the sigmoid function significantly reduces the time cost.

As is demonstrated in \ref{sec:appendix_numerical_method}, to obtain the steady state, we are essentially evolving the kinetic equation for long enough time. Therefore the time cost is inversely proportional to the time step size $\Delta t$ we use. Recall that the marginality spectrum $R(\mathbf{k})$ depends on electron distribution $f$ linearly, therefore within one time step, $\delta R \propto \delta f = \mathcal{O}(\Delta t)$. Moreover, considering that near marginality, $R\sim 1$, we require that $\delta R \ll 1-R$, which renders an upper bound for $\Delta t$. To balance accuracy and efficiency, we expect $R \sim 0.9$ under our choice of parameters $\sigma$ or $(W_{max}, \alpha)$.

\bigskip
For a small enough parameter $\sigma$, suppose that the associated steady state is $f$, and the marginality spectrum is $R(\mathbf{k})$, which corresponds to a wave spectrum $W(\mathbf{k})$. 

Fix $k_{\parallel}$ and set $z=k_{\perp}^{2}$. Near the maximum of the marginality spectrum, assume 
\begin{equation*}
    R(z) \approx (1-\varepsilon) - \lambda (z-B^{2})^{2}, \qquad \varepsilon = 1-R(B^{2}) > 0,
\end{equation*}
then $\varepsilon$ measures the distance from exact marginality.

\begin{itemize}
    \item Under the original wave response \eqref{eq:orginalW}, the amplitude $A$ at $k_{\parallel}$ satisfies
\begin{equation*}
    A = \int_{0}^{\infty} \frac{\sigma}{1-\left[(1-\varepsilon) - \lambda (z-B^{2})^{2}\right]} \frac{1}{2B} dz \approx \frac{\sigma}{2B}\frac{\pi}{\sqrt{\varepsilon \lambda}}.
\end{equation*}
It follows that
\begin{equation}\label{eq:ep_original}
    \varepsilon = \frac{\pi^{2}\sigma^{2}}{4 B^{2}A^{2}\lambda}.
\end{equation}

\item For the sigmoid wave response, the same derivation gives
\begin{equation*}
    A = \int_{0}^{\infty} \frac{W_{max}}{1 + \exp \left[-\alpha((1-\varepsilon)-\lambda(z-B^{2})^{2}-1)\right]}\frac{1}{2B}dz \approx \frac{W_{max}}{2B}\exp(-\alpha \varepsilon) \frac{\sqrt{\pi}}{\sqrt{\alpha \lambda}},
\end{equation*}
which yields the following relation
\begin{equation}\label{eq:ep_sigmoid}
    \varepsilon = \frac{1}{2\alpha} \ln\left(\frac{\pi W_{max}^{2}}{4 \alpha B^{2}A^{2} \lambda}\right).
\end{equation}
\end{itemize}

Note that the amplitude $A$ and the maximum point $B$ are determined by the spectrum $W$, moreover the coefficient $\lambda$ related to the shape of marginality spectrum $R$ is determined by the electron distribution $f$ at steady state. In other words, $A$, $B$ and $\lambda$ are given functions which varies with $k_{\parallel}$.

Comparing \eqref{eq:ep_original} and \eqref{eq:ep_sigmoid}, it can be observed that the sigmoid wave response compresses the variation of epsilon across $k_{\parallel}$ modes and prevents a small subset of modes from imposing an excessively small time step.

\section{Numerical experiments}\label{sec:experiment}

The numerical experiment results presented in this paper are all performed under the following set of parameters: Coulomb logarithm $\ln{\Lambda} = 10$, electron number density $n_{e} = 1.0 \times  10^{20} m^{-3}$, electron gyro-frequency $\omega_{ce} = -0.4 \omega_{pe}$, charge of impurity $Z = 6$, and knock-on collision cut-off parameter $\gamma_{m} = 1.4$.

\subsection{Three regimes at fixed bulk temperature}
In this subsection, we will show how the quasi-steady RE distribution depends on the total current density, under a fixed electron bulk temperature.

We denote the avalanche threshold field by $E_{av}$. At fixed $T_e$, this field defines the avalanche onset current $j_0=E_{av}/\eta$. The second threshold, denoted by $j_1$, is the total current density at which the runaway-electron population first drives the whistler instability before the inductive field can relax back to $E_{av}$.

\begin{figure}[H]
    \centering
    \includegraphics[width=0.75\textwidth]{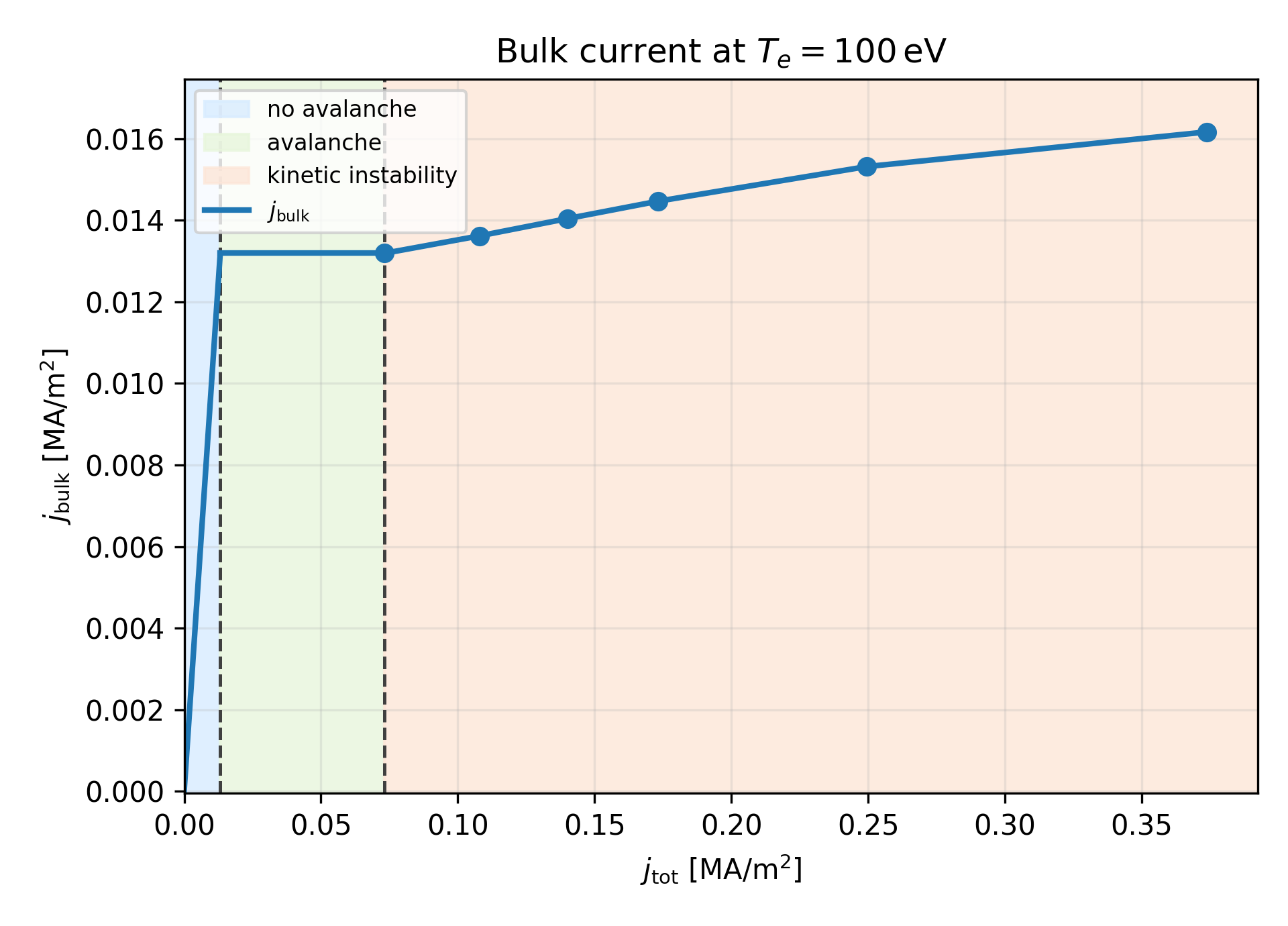}
    \caption{Three phases at fixed bulk temperature $T_{e} = 100 eV$, illustrated by the final bulk current density as a function of total current density.}
    \label{fig:phases_varyingItot}
\end{figure}

Figure \ref{fig:phases_varyingItot} shows the relation between bulk current and total current at the quasi-steady state. As the parameter $j_{tot}$ is swept over $(0, \infty)$, the system exhibits three qualitatively distinct phases, separated by transition points at avalanche onset current $j_{0}$ and instability onset current $j_{1}$:
\begin{itemize}
    \item If $j_{tot}$ is less than the avalanche onset current $j_{0}$, initially there is no runaway current at all, and the induced field $E = \eta j_{bulk} = \eta j_{tot}$ is below the avalanche threshold $E_{av}$. Therefore all the runaway electrons will return to the bulk eventually, i.e. $j_{bulk} = j_{tot}$ and $j_{RE}=0$;
    \item If $j_{tot}$ is between the avalanche onset current $j_{0}$ and instability onset current $j_{1}$, at first the induced electric field is above the avalanche threshold, and at this point the RE population grows exponentially. The bulk current decreases as more and more runaway electrons are created through knock-on collisions, hence the induced electric field will decay. The decay of electric field will not stop until it reaches the avalanche threshold, which is $E_{av}$, therefore the system ends up with $j_{bulk} = E_{av}/\eta$ and $j_{RE} = j_{tot} - E_{av}/\eta$;
    \item If $j_{tot}$ is above the instability onset current $j_{1}$, in the beginning, the RE population grows exponentially. However, before the RE current reaches $j_{tot} - E_{av}/\eta$, the excitation due to non-Maxwellian RE distribution overcomes the collisional damping, and the kinetic instability will be triggered. The growing whistler spectrum enhances momentum-space diffusion until diffusive losses balance knock-on production, so the avalanche growth saturates. As a consequence, eventually the RE current density is below $j_{tot} - E_{av}/\eta$, and the bulk current density is above $E_{av}/\eta$.
\end{itemize}

\begin{figure}[H]
    \centering
    \includegraphics[width=0.75\textwidth]{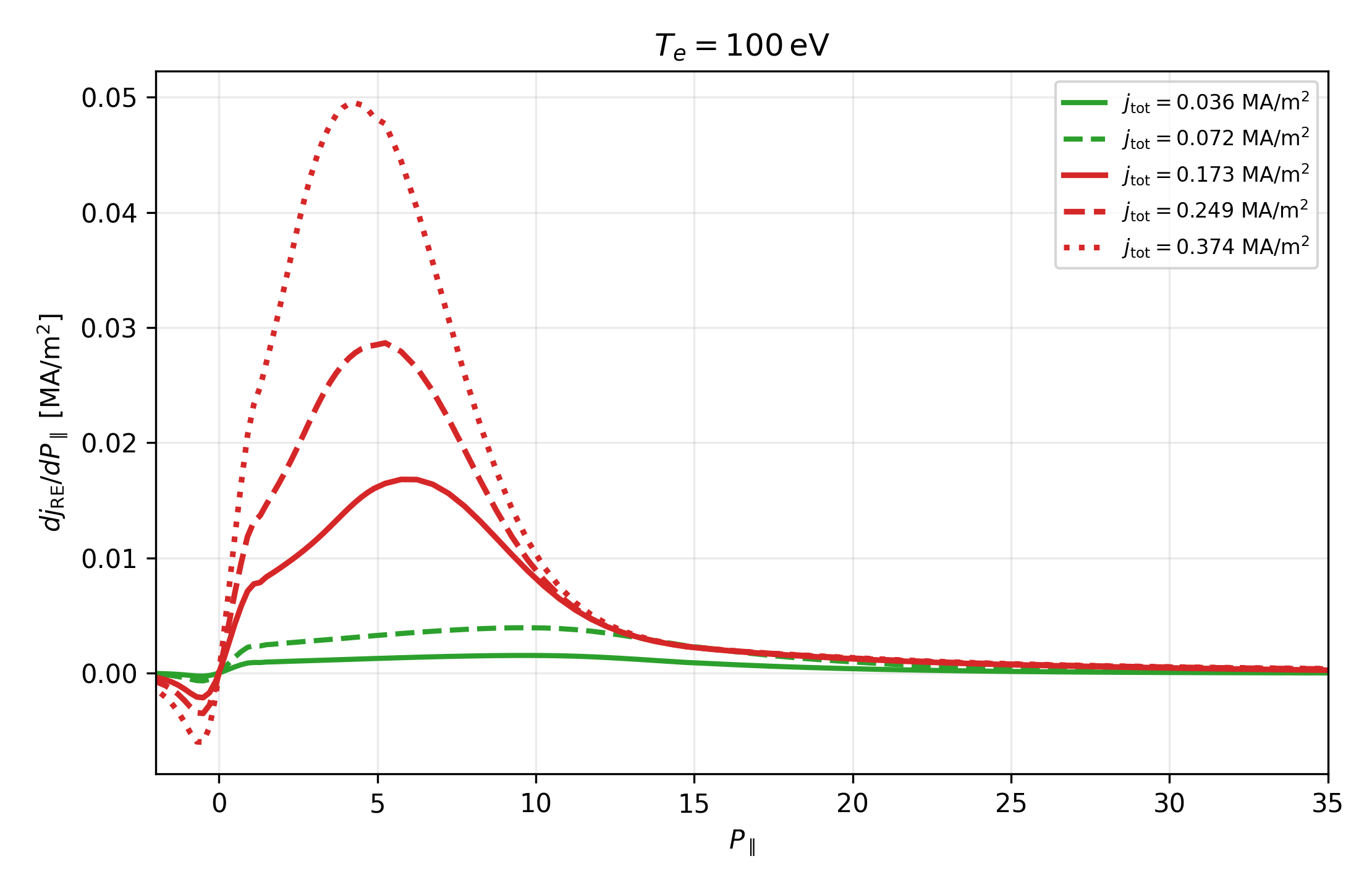}
    \caption{RE current distribution $d j_{\mathrm{RE}}/dP_{\parallel}$ under different total currents at fixed bulk temperature $T_e=100\,\mathrm{eV}$. Green curves correspond to phase-II examples, while red curves correspond to phase-III examples at the same prescribed bulk temperature. In phase II, varying $j_{tot}$ mainly rescales the distribution. In phase III, self-excited whistler waves change the distribution shape; nevertheless, the high-energy tails for $P_{\parallel}>15$ nearly coincide.}
    \label{fig:RE_distribution_varyingItot}
\end{figure}

In Figure \ref{fig:RE_distribution_varyingItot} we plot the RE current distribution $\frac{d j_{\mathrm{RE}}}{dP_{\parallel}} (P_{\parallel})$ under various total currents:
\begin{equation*}
    \frac{d j_{\mathrm{RE}}}{dP_{\parallel}} (P_{\parallel})
    =
    m^{3}c^{4}\int f(P_{\parallel}, P_{\perp})\frac{P_{\parallel}}{\gamma}
    2\pi P_{\perp}\,dP_{\perp}.
\end{equation*}
The two green curves are examples in phase-II: avalanche regime, where increasing $j_{tot}$ mainly rescales the phase-II distribution. The red curves are in phase-III: instability-regulated regime. They show two key features. First, once $j_{tot}>j_1$, the phase-III distribution cannot be obtained by simply rescaling a phase-II distribution; the onset of self-excited waves changes the shape of the distribution. Second, the high-energy tail, roughly $P_\parallel = \frac{p_{\parallel}}{mc}>15$, is nearly universal across the phase-III cases. This universality indicates that the tail is controlled primarily by the excitation-damping balance of whistler waves, rather than by the precise value of $j_{tot}$.

At the steady state, the time derivative of any macroscopic quantity is equal to zero, i.e. $\frac{d}{dt}\left(\int_{\Omega^{*}}f\phi d^{3}\mathbf{p}\right)=0,\ \forall \phi$, because the contributions from various processes balance with each other. Since all the terms except the knock-on source are in divergence form, their contributions can be calculated as follows:
\begin{equation*}
    \int_{\Omega_{*}} (\nabla_{p}\cdot X) \phi d^{3}\mathbf{p} = - \int_{\Omega_{*}} X \cdot \nabla_{p} \phi d^{3}\mathbf{p} + \oint_{\partial \Omega_{*}} \phi X\cdot\mathbf{n} dS,\qquad \Omega_{*}\coloneqq\{\mathbf{p}:p_{\parallel}>mc\}.
\end{equation*}

\begin{figure}[htbp]
    \centering
    \includegraphics[width=0.75\textwidth]{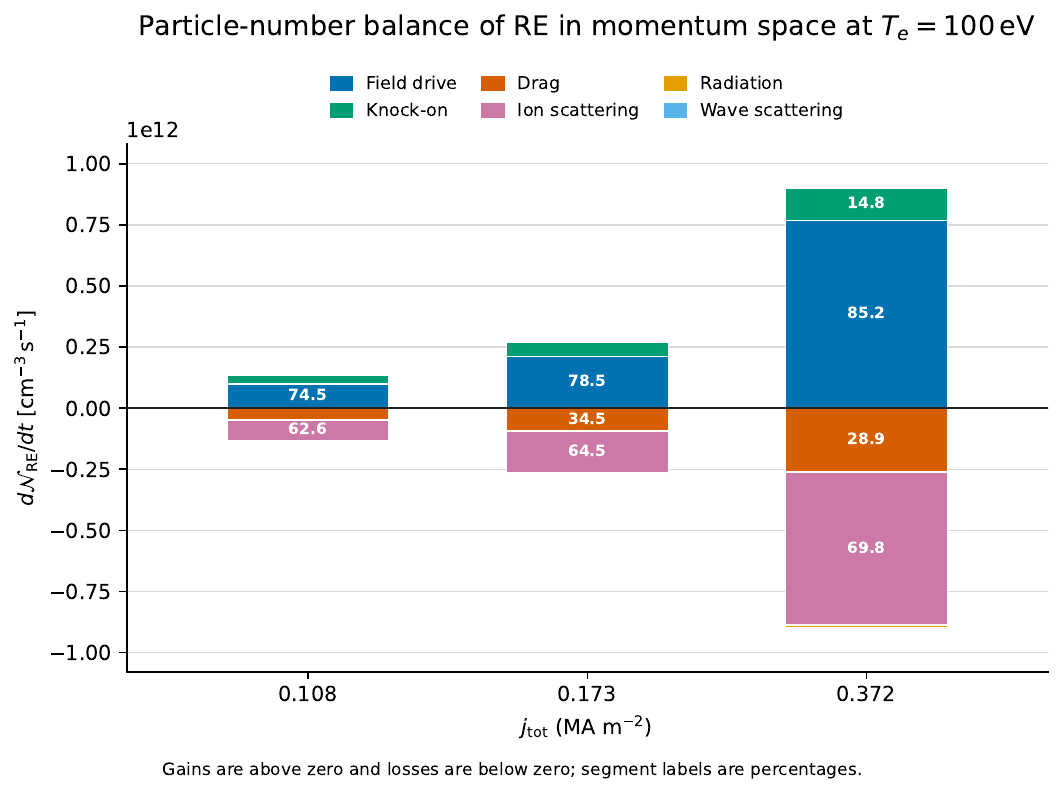}
    \caption{Contributions of various processes to the particle-number balance of RE at $T_e=100\,\mathrm{eV}$. $\mathcal{N}_{RE} = \int_{p_{\parallel}>mc}fd^{3}\mathbf{p}$.}
    \label{fig:RE_mass_transfer}
\end{figure}

\begin{figure}[htbp]
    \centering
    \includegraphics[width=0.75\textwidth]{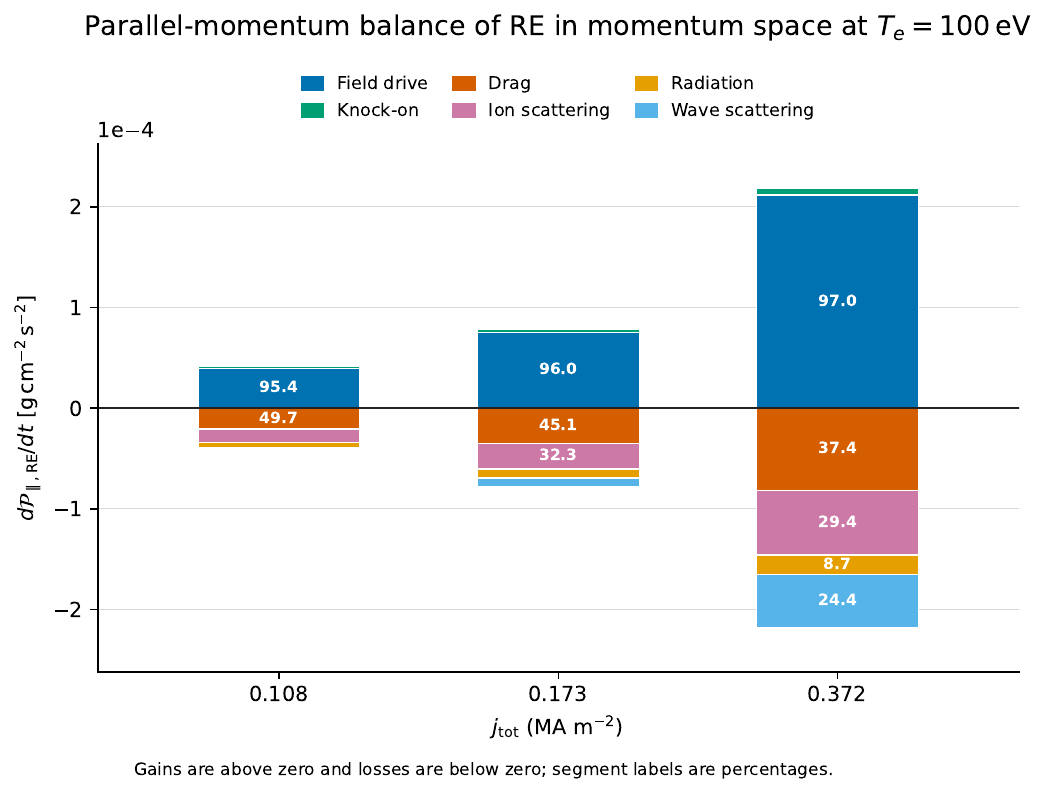}
    \caption{Contributions of various processes to the momentum balance of RE at $T_e=100\,\mathrm{eV}$. $\mathcal{P}_{\parallel,RE} = \int_{p_{\parallel}>mc}f p_{\parallel}d^{3}\mathbf{p}$.}
    \label{fig:RE_momentum_transfer}
\end{figure}

\begin{figure}[htbp]
    \centering
    \includegraphics[width=0.75\textwidth]{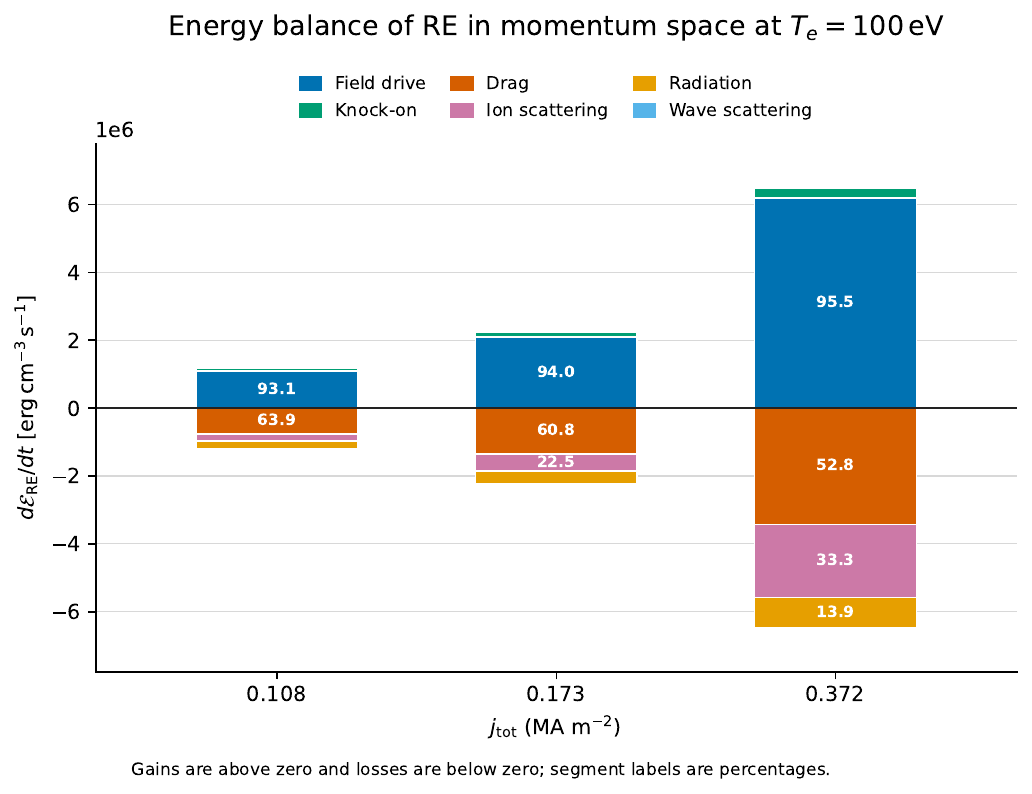}
    \caption{Contributions of various processes to the energy balance of RE at $T_e=100\,\mathrm{eV}$. $\mathcal{E}_{RE} = \int_{p_{\parallel}>mc}f \sqrt{1+|\mathbf{p}|^{2}}d^{3}\mathbf{p}$.}
    \label{fig:RE_energy_transfer}
\end{figure}

Consider the sub-domain where $p_{\parallel}>mc$, Figures \ref{fig:RE_mass_transfer}--\ref{fig:RE_energy_transfer} show the balance of mass, momentum and energy.

\begin{remark}
    Note that all of those contributions consist of both a volume integral part and a boundary integral part. For example, although the ion scattering does not change the energy of a single electron, it can contribute to energy loss through an outward flux on the boundary.
\end{remark}

\subsection{Phase diagram in total current and bulk temperature}

\begin{figure}[H]
    \centering
    \begin{subfigure}[b]{0.48\textwidth}
        \centering
        \includegraphics[width=\textwidth]{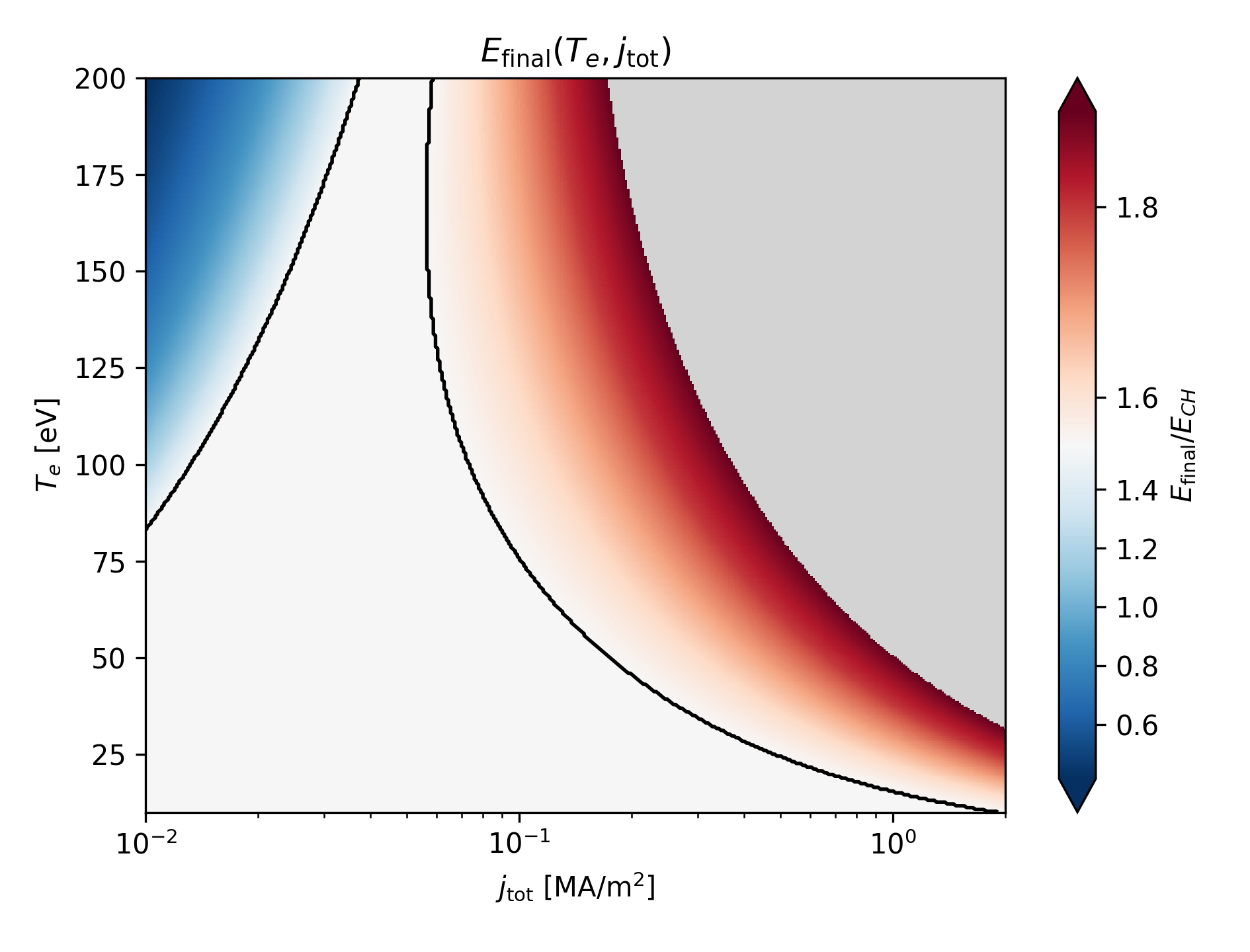}
        \caption{Final electric field}
    \end{subfigure}
    \hfill
    \begin{subfigure}[b]{0.48\textwidth}
        \centering
        \includegraphics[width=\textwidth]{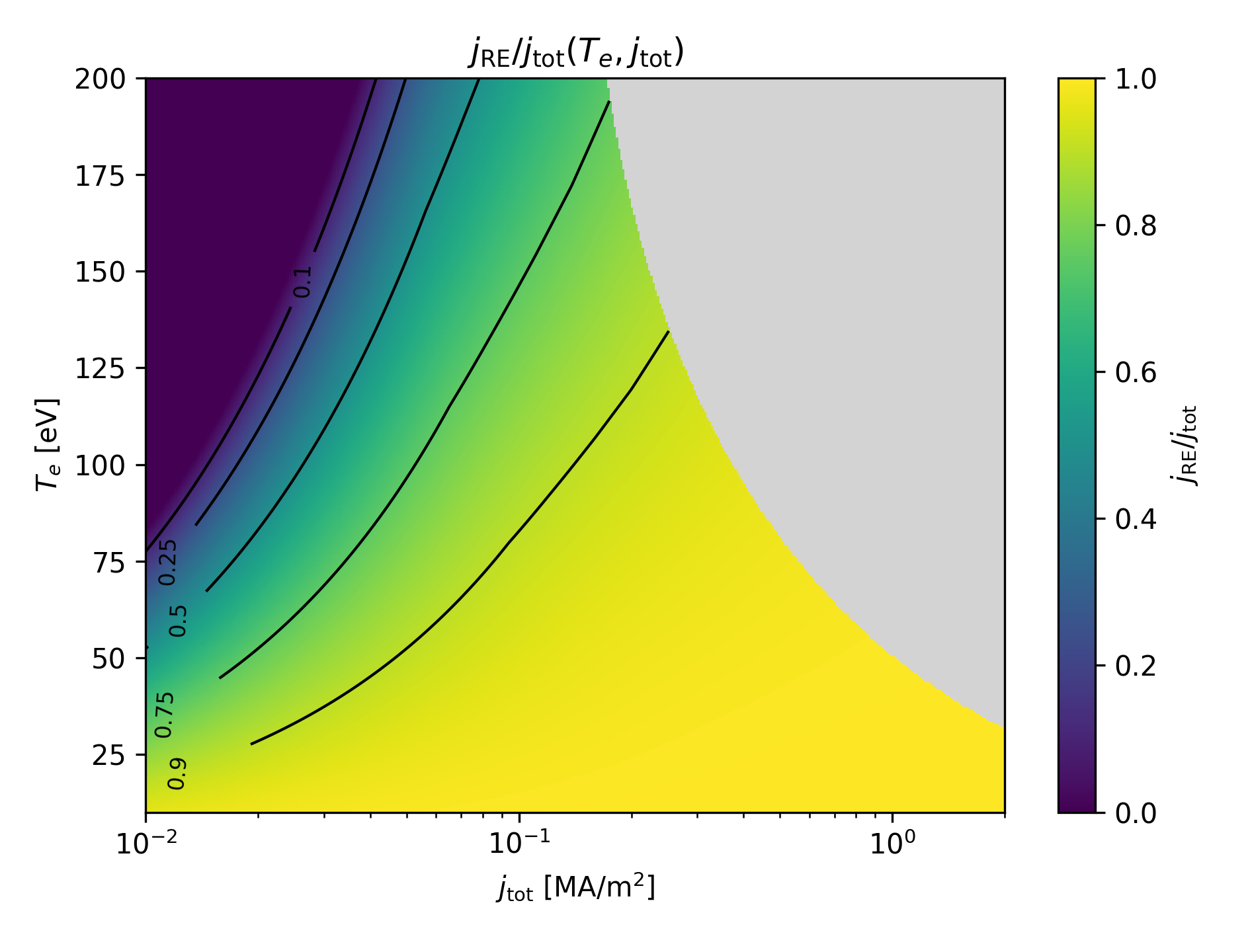}
        \caption{RE current fraction}
    \end{subfigure}
    \caption{Phase diagram in the $(j_{tot},T_e)$ parameter plane. (The gray region marks parameter values outside the finite set of numerical experiments; it is not a physical boundary and does not indicate a limitation of the kinetic model or numerical method.)}
    \label{fig:phase_diagram_jtot_T}
\end{figure}

Figure \ref{fig:phase_diagram_jtot_T} summarizes the quasi-steady state over the parameter plane spanned by total current density and bulk temperature. The final electric field identifies the three regimes discussed above, where the blue region represent the no-avalanche phase, the red region correspond to the instability-regulated regime, and the white region between them is the avalanche phase. The RE current fraction shows how the total current is partitioned between the bulk and runaway-electron components.

\subsection{RE distribution at various bulk temperature}
At fixed total current density $j_{tot} = 0.16\,\mathrm{MA}/\mathrm{m}^{2}$, we compare the quasi-steady RE current distributions obtained at different prescribed bulk temperatures. Along a parameter scan in which $T_e$ is varied from high to low values, the RE population first increases. After the RE current nearly saturates, further lowering $T_e$ mainly shifts the distribution toward higher parallel momentum. For sufficiently low $T_e$, however, collisional damping is strong enough that the kinetic instability is not triggered.

\begin{figure}[H]
    \centering
    \includegraphics[width=0.75\textwidth]{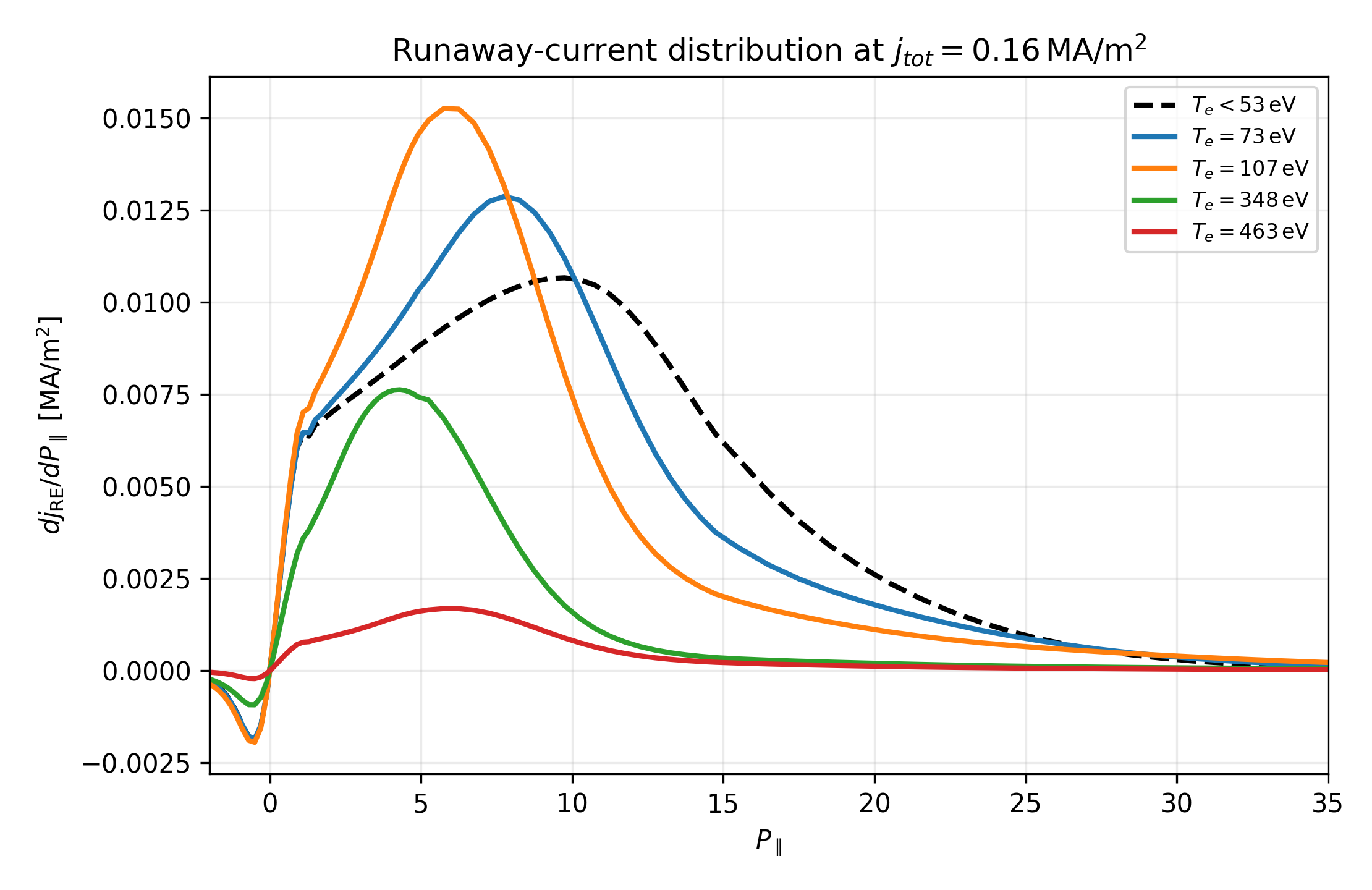}
    \caption{RE current distribution under different bulk temperature.}
    \label{fig:RE_distribution_varyingTbulk}
\end{figure}

\subsection{Spectrum of the whistler waves}

\begin{figure}[htbp]
    \centering
    \begin{subfigure}[b]{0.48\textwidth}
        \centering
        \includegraphics[width=\textwidth]{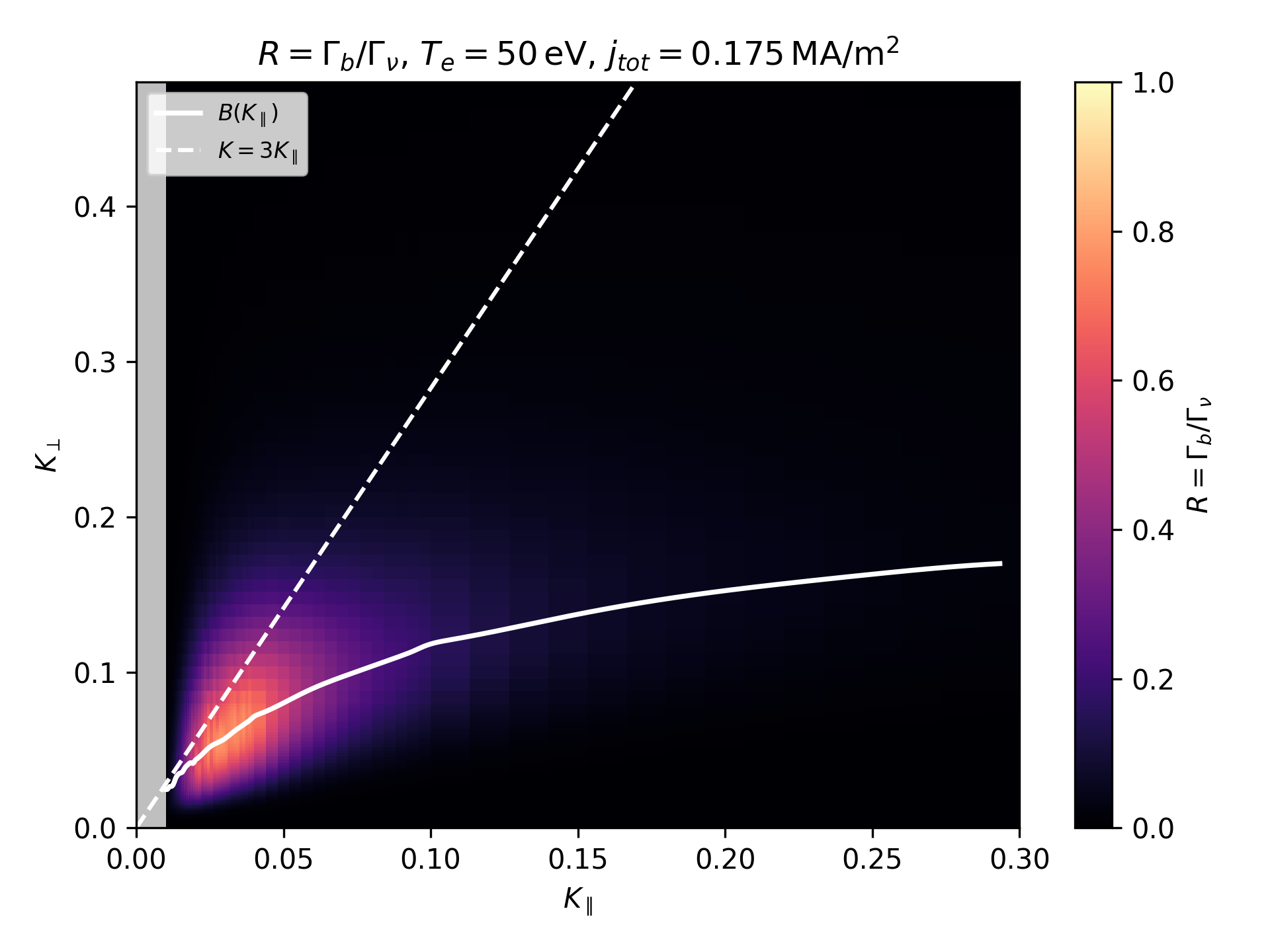}
        \caption{$R(\mathbf{k})$, $j_{tot} =0.175\,\mathrm{MA}/\mathrm{m}^{2}$}
    \end{subfigure}
    \hfill
    \begin{subfigure}[b]{0.48\textwidth}
        \centering
        \includegraphics[width=\textwidth]{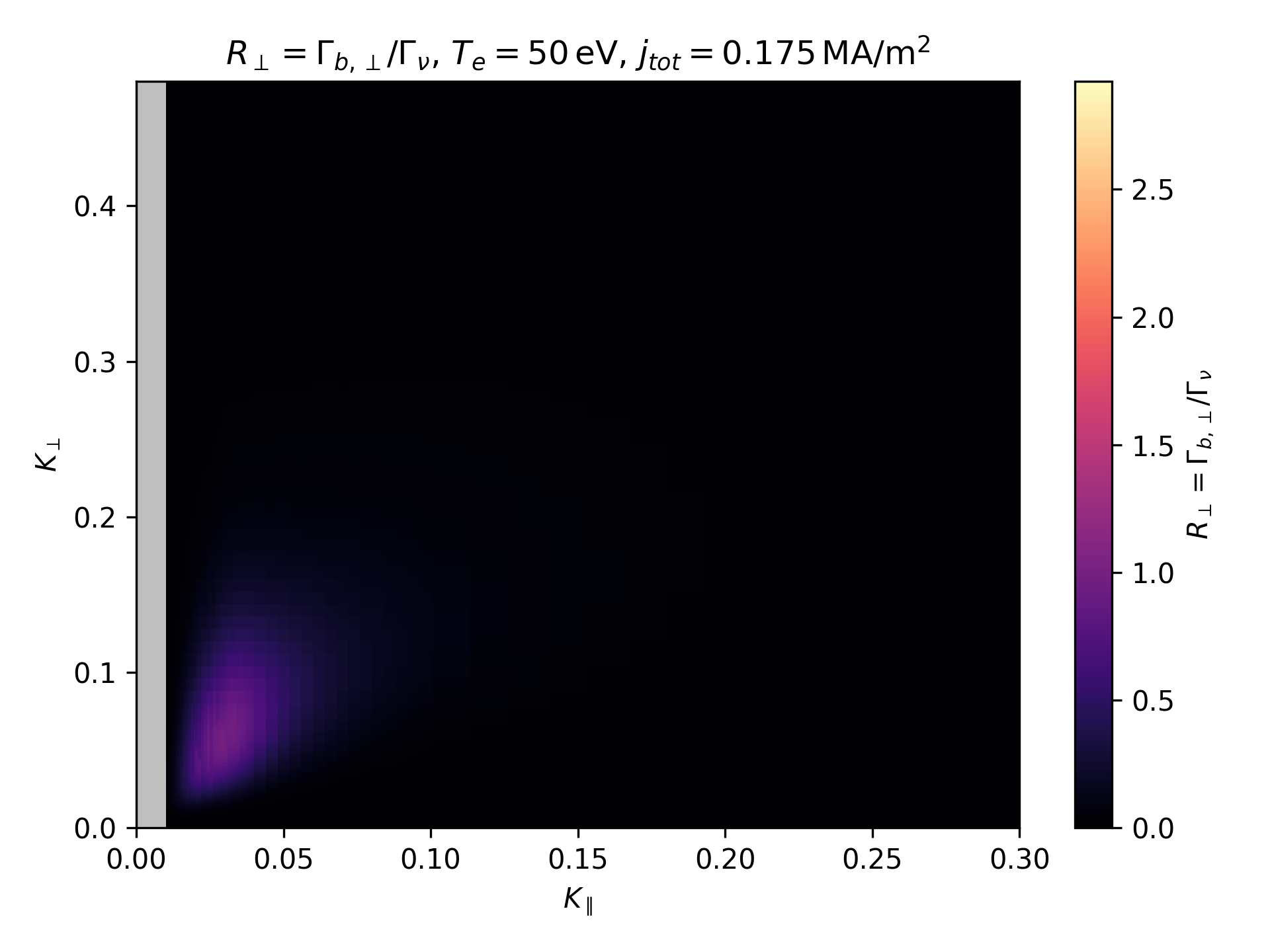}
        \caption{$R_\perp(\mathbf{k})$, $j_{tot} =0.175\,\mathrm{MA}/\mathrm{m}^{2}$}
    \end{subfigure}
    \begin{subfigure}[b]{0.48\textwidth}
        \centering
        \includegraphics[width=\textwidth]{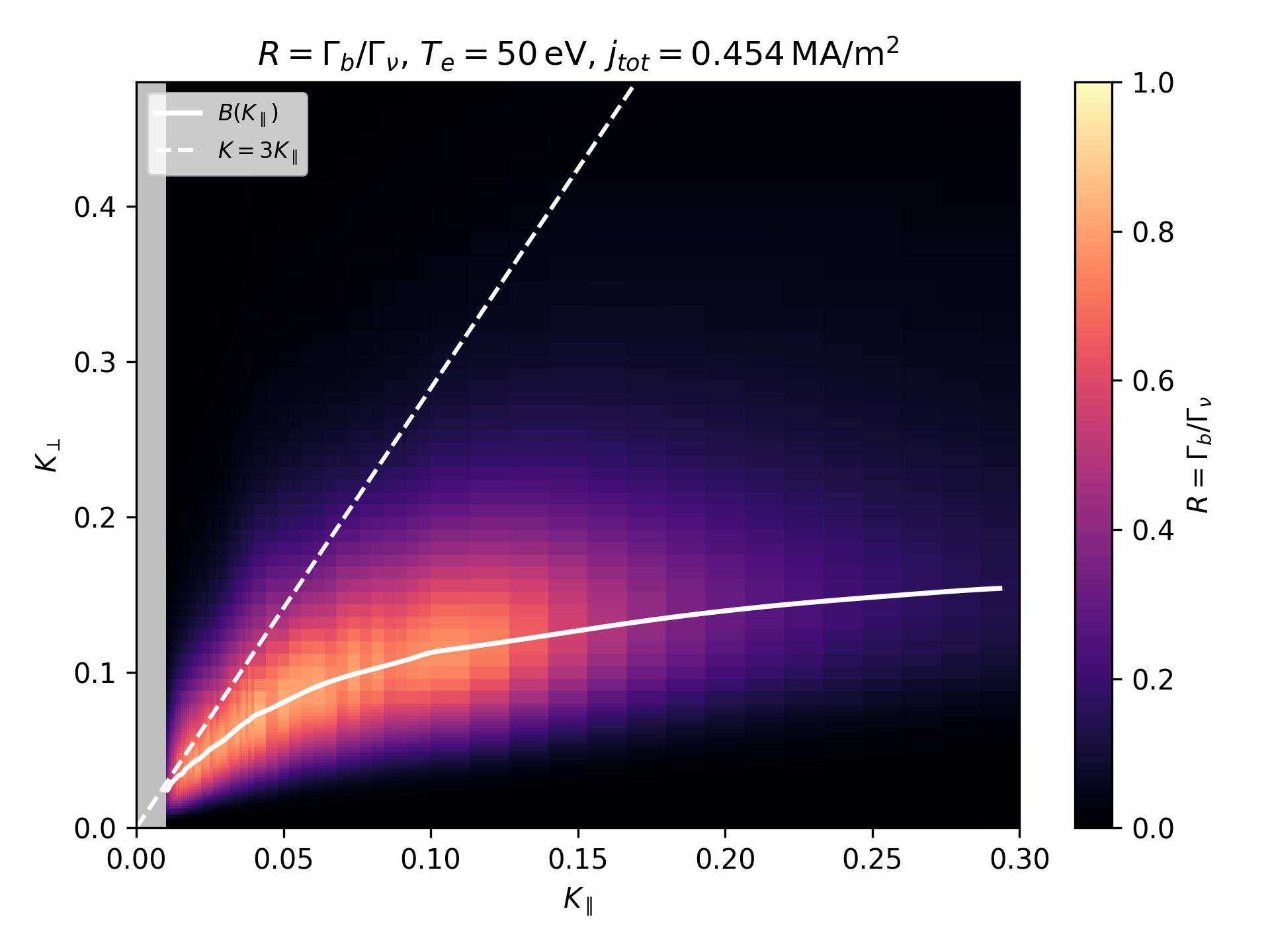}
        \caption{$R(\mathbf{k})$, $j_{tot} =0.454\,\mathrm{MA}/\mathrm{m}^{2}$}
    \end{subfigure}
    \hfill
    \begin{subfigure}[b]{0.48\textwidth}
        \centering
        \includegraphics[width=\textwidth]{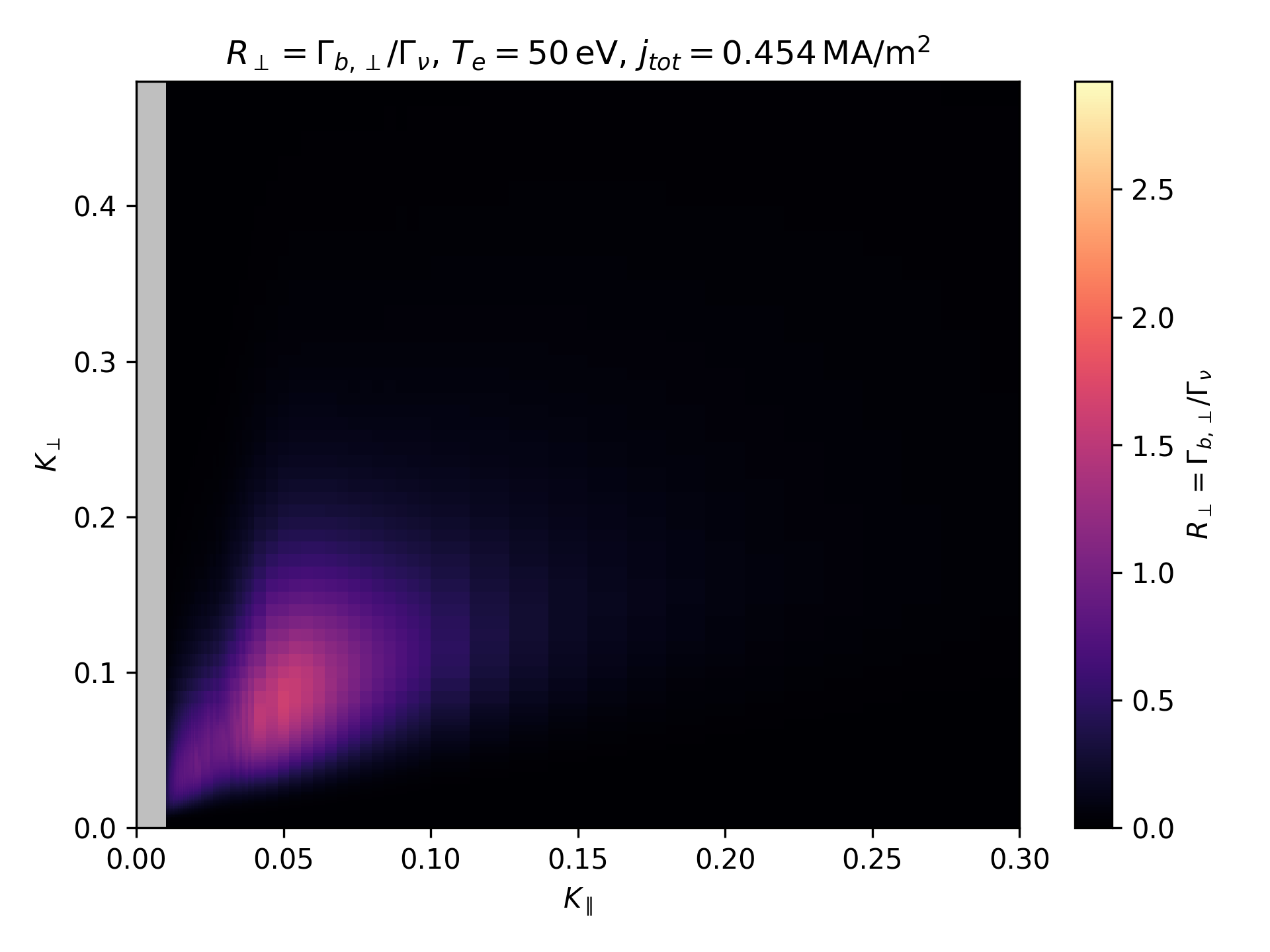}
        \caption{$R_\perp(\mathbf{k})$, $j_{tot} =0.454\,\mathrm{MA}/\mathrm{m}^{2}$}
    \end{subfigure}
    \begin{subfigure}[b]{0.48\textwidth}
        \centering
        \includegraphics[width=\textwidth]{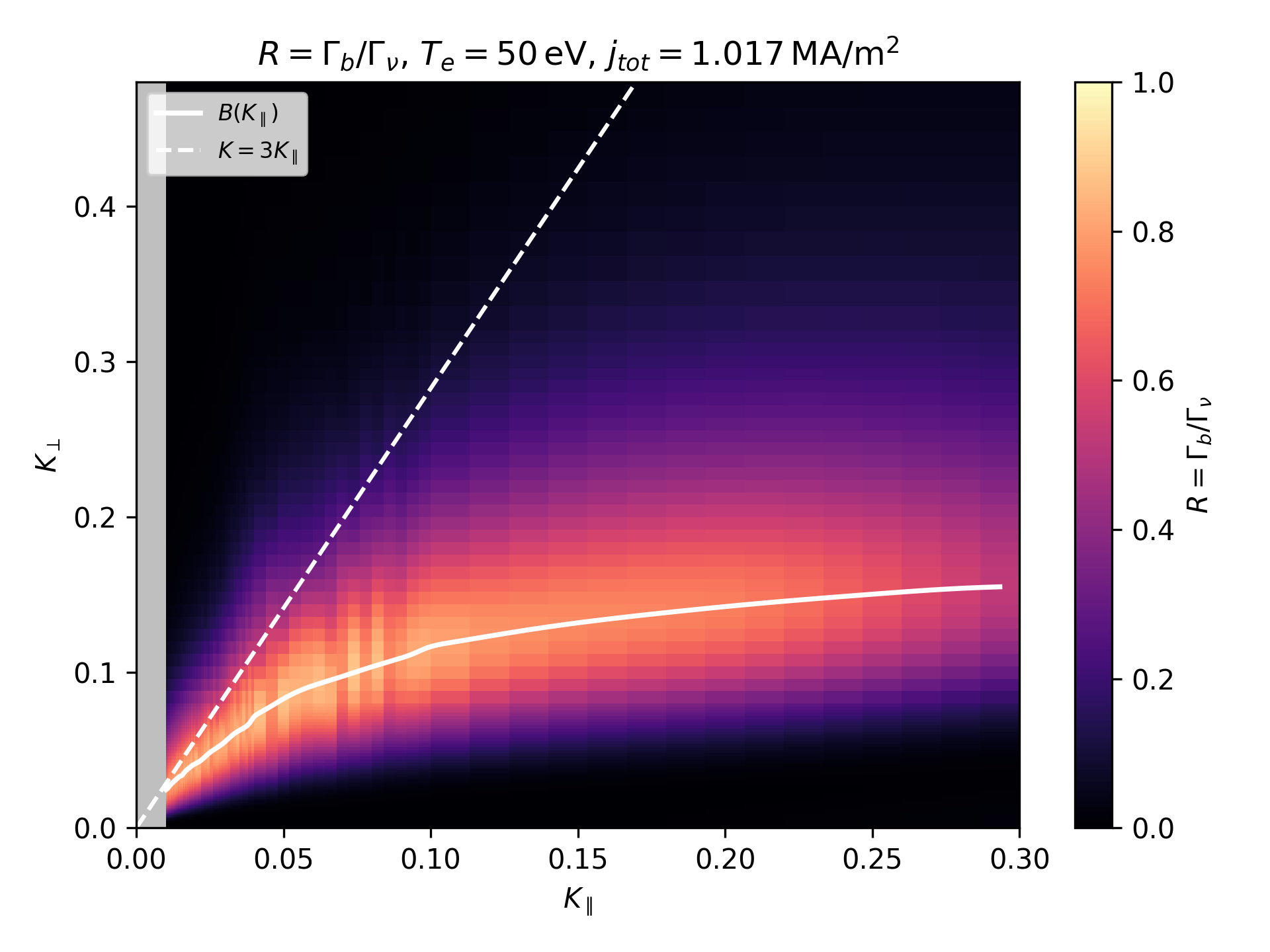}
        \caption{$R(\mathbf{k})$, $j_{tot} =1.017\,\mathrm{MA}/\mathrm{m}^{2}$}
    \end{subfigure}
    \hfill
    \begin{subfigure}[b]{0.48\textwidth}
        \centering
        \includegraphics[width=\textwidth]{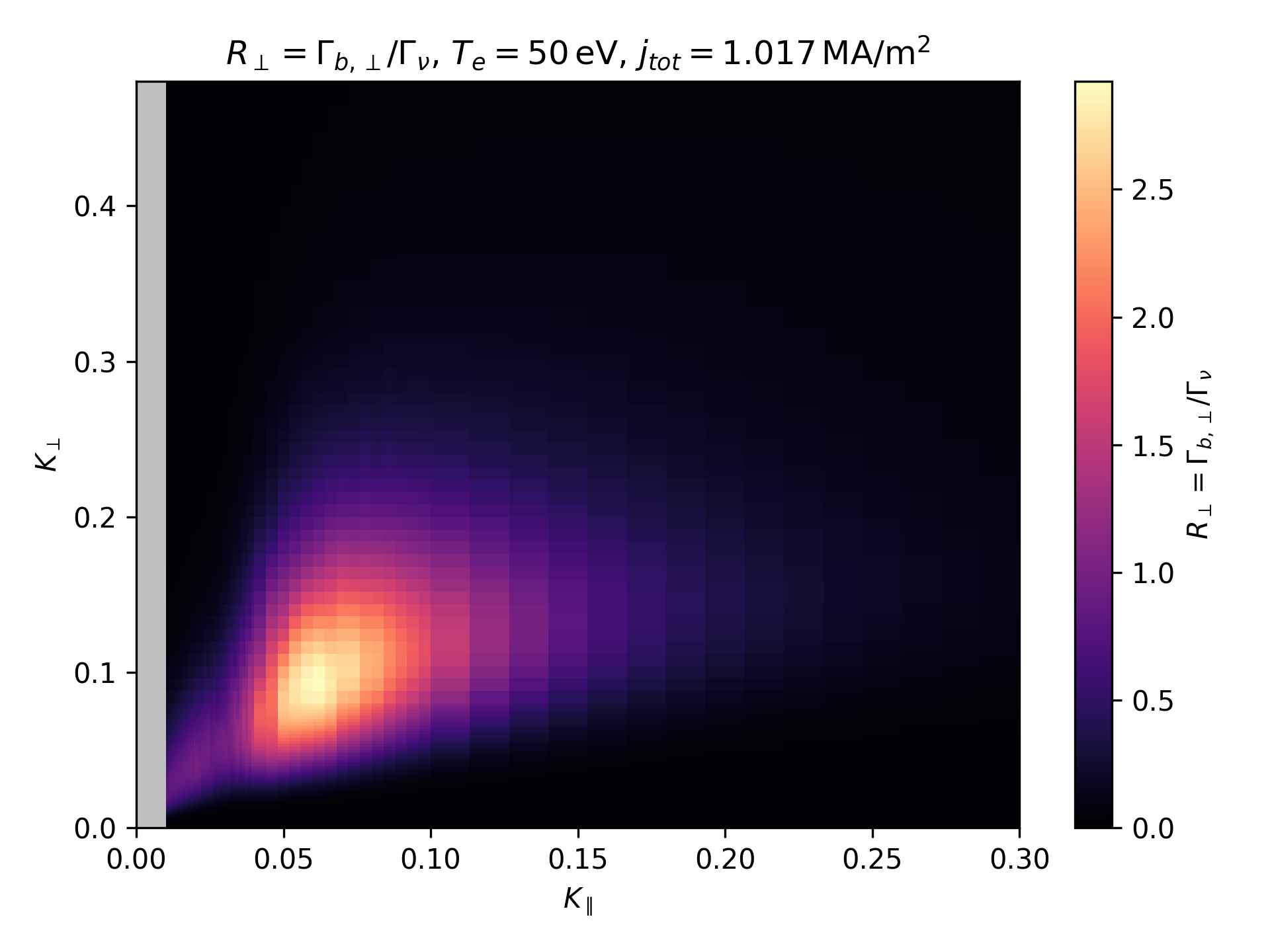}
        \caption{$R_\perp(\mathbf{k})$, $j_{tot} =1.017\,\mathrm{MA}/\mathrm{m}^{2}$}
    \end{subfigure}
    \caption{Excitation-to-damping ratio $R=\Gamma_b/\Gamma_\nu$ and the associated perpendicular-gradient contribution $R_\perp=\Gamma_{b,\perp}/\Gamma_\nu$ at $T_e=50\,\mathrm{eV}$ for three corrected total-current densities. The axes use $\mathbf{K}=\mathbf{k}c/\omega_p$. The left column shows $R$ with the fitted ridge $B(K_\parallel)$ and the small-$k$ reference $k_{\perp}=2\sqrt{2}k_\parallel$ overlaid in white. The comparison shows that the perpendicular-gradient drive alone cannot explain the wave spectrum at larger $K_{\parallel}$, especially for $K_{\parallel}  > 0.05$.}
    \label{fig:spectrum}
\end{figure}

In the modeling part, we have assumed that the wave spectrum can be approximated with a Dirac delta in $k_{\perp}$ direction. This assumption is justified a posteriori through numerical experiments, see Figure \ref{fig:spectrum}. 

In \eqref{eq:magreaction}, $\beta\cdot\nabla_p f$ contains both the parallel-gradient and perpendicular-gradient contributions to the wave drive. Below we denote the contribution from the perpendicular-gradient part by $\Gamma_{b,\perp}$ when diagnosing the numerical results, i.e.
\begin{equation*}
    \Gamma_{b,\perp}(\mathbf{k},t)= \frac{1}{2} \sum_{l}\int d^{3} \mathbf{p} \left((1-\frac{k_{\parallel}v_{\parallel}}{\omega})\frac{p}{p_{\perp}}\frac{\partial f}{\partial p_{\perp}}\right)\left(\beta \cdot \nabla_{p}\gamma mc\right)U_{l}(\mathbf{p},\mathbf{k})\delta(\omega-k_{\parallel}v_{\parallel}-l\omega_{c}/\gamma).
\end{equation*}

Figure \ref{fig:spectrum} shows two wave diagnostics at fixed bulk temperature $T_e=50\,\mathrm{eV}$ for three values of the total current:
\begin{equation*}
    R(\mathbf{k})=\frac{\Gamma_b(\mathbf{k})}{\Gamma_\nu(\mathbf{k})},\qquad
    R_\perp(\mathbf{k})=\frac{\Gamma_{b,\perp}(\mathbf{k})}{\Gamma_\nu(\mathbf{k})}.
\end{equation*}
The plotted wave-number axes use the dimensionless variable
\begin{equation*}
    \mathbf{K}\equiv \frac{\mathbf{k}c}{\omega_p},\qquad
    K_\parallel=\frac{k_\parallel c}{\omega_p},\qquad
    K_\perp=\frac{k_\perp c}{\omega_p}.
\end{equation*}

In the left-column panels, the solid white curve marks the fitted ridge location $B(K_\parallel)$ used in the one-dimensional delta-spectrum approximation, and the dashed white line is the small-$k$ theoretical reference \cite{breizman2023marginal} $k=\sqrt{k_\parallel^2+k_\perp^2}=3k_\parallel$. The computed domain begins at a finite $k_{\parallel,\min}$, and the gray strip indicates the excluded region between $k_\parallel=0$ and the first computed cell.

The numerical ridge follows the theoretical small-$k$ trend at low $k_\parallel$, where the resonance is dominated by particles with $p_\parallel\gg p_\perp$. At larger $k_\parallel$, the ridge bends away from this asymptotic reference, reflecting the finite-width runaway distribution and the full two-dimensional resonance geometry.

\begin{remark}\label{remark:Alfven}
The spectra in Figure \ref{fig:spectrum} are computed with the whistler dispersion relation. This is an important simplification. The full cold-plasma dielectric tensor and the definitions of $\omega_{ps}$ and $\omega_{cs}$ are given in Appendix \ref{sec:appendix_kinetic_instability}. In those terms, the whistler approximation keeps the electron response while neglecting ion motion. This is accurate only when the frequency and parallel wavelength make the ion terms small. At sufficiently small $K_\parallel$, however, the wave frequency approaches the ion-response range, ion motion is no longer negligible, and the whistler branch connects to the compressional Alfv\'en mode. In this paper we present the results obtained with the whistler dispersion relation for simplicity.
\end{remark}

\section{Conclusion}\label{sec:conclusion}

We have formulated a reduced description of the long-time runaway-electron state in which the total plasma current is treated as the control parameter and the inductive electric field is determined self-consistently by the partition between bulk and runaway current. This formulation separates the fast kinetic relaxation from the slower current-decay dynamics and allows the quasi-steady runaway distribution to be obtained from constant-field numerical experiments.

The resulting states fall into three regimes. Below the avalanche-onset current, runaway electrons decay and the current remains Ohmic. Between the avalanche-onset and instability-onset currents, the avalanche saturates by reducing the inductive electric field to the avalanche threshold. Above the instability-onset current, the growing whistler waves enhances momentum-space diffusion until diffusive losses balance knock-on
production, so the avalanche growth saturates before the runaway current reaches the phase-II value. The numerical scans show how these regimes organize the parameter plane spanned by total current density and bulk temperature.

The computed excitation-to-damping ratio also supports the one-dimensional approximation to the excited wave spectrum. In the unstable regime, the near-marginal region forms a narrow ridge in $(k_\parallel,k_\perp)$ space, and the ridge follows the expected small-$k$ theoretical trend before bending away at larger $k_\parallel$. This provides an a posteriori justification for replacing the saturated spectrum by a delta-like distribution in $k_\perp$ when constructing the quasilinear diffusion operator.

These results suggest that wave-regulated marginal stability can set a robust upper bound on the runaway current at fixed total current and temperature. In addition, we find that the runaway current is carried largely by electrons at modest energies, rather than by the high-energy tail of the distribution, which has also been observed in experiments \cite{hollmann2015measurement}.

A natural next step is to remove the prescribed bulk temperature as an independent control parameter. In the present work we used $(j_{tot},T_e)$ as a two-parameter description to map the marginal state selected by the balance between runaway-driven whistler growth and collisional wave damping. In an experiment, however, the external control is closer to the total current evolution, while $T_e$ is itself determined by the coupled post-disruption energy balance. As emphasized in \cite{breizman2019physics}, the bulk temperature during the thermal and current quench is shaped by impurity radiation, Ohmic heating, density evolution, ionization, magnetic transport, and the energy transferred to or from the runaway population. A more predictive model should therefore couple the present marginal-stability closure to an evolution equation for the bulk plasma temperature and density, so that the runaway current, inductive electric field, wave-regulated distribution, and collisional damping are determined self-consistently during the current decay.

\section*{Acknowledgements}
This work was supported by the US DOE under Grant DE-FG02-04ER54742

\appendix
\section{Kinetic instability}\label{sec:appendix_kinetic_instability}

We use the cold-plasma dielectric tensor in the Hermitian limit. Let $\mathbf{b}=\mathbf{B}/B$ be the unit vector along the background magnetic field and let $\epsilon_{\alpha\beta\gamma}$ denote the Levi-Civita tensor. The tensor is
\begin{equation*}
    \varepsilon^H_{\alpha\beta}
    =
    \varepsilon\left(\delta_{\alpha\beta}-b_\alpha b_\beta\right)
    +\eta b_\alpha b_\beta
    + i g\,\epsilon_{\alpha\beta\gamma}b_\gamma,
\end{equation*}
with scalar components
\begin{equation*}
    \varepsilon = 1-\sum_s\frac{\omega_{ps}^{2}}{\omega^{2}-\omega_{cs}^{2}},
    \qquad
    g = -\sum_s\frac{\omega_{cs}}{\omega}
    \frac{\omega_{ps}^{2}}{\omega^{2}-\omega_{cs}^{2}},
    \qquad
    \eta = 1-\sum_s\frac{\omega_{ps}^{2}}{\omega^{2}}.
\end{equation*}
Here the sum is over particle species, $\omega_{ps}^{2}=4\pi n_s q_s^2/m_s$ is the plasma frequency squared, and $\omega_{cs}=q_sB/(m_sc)$ is the signed cyclotron frequency. The wave polarization vector is obtained from the cold-plasma wave equation
\begin{equation*}
    \left(N_\alpha N_\beta-\delta_{\alpha\beta}N^2+\varepsilon^H_{\alpha\beta}\right)E_\beta=0,
    \qquad
    \mathbf{N}=\frac{c\mathbf{k}}{\omega}.
\end{equation*}
Writing $N_\parallel=\mathbf{N}\cdot\mathbf{b}$, $N_\perp=\sqrt{N^2-N_\parallel^2}$, $\mathbf{e}_1=(\mathbf{N}-N_\parallel\mathbf{b})/N_\perp$, and $\mathbf{e}_2=\mathbf{b}\times\mathbf{N}/N_\perp$, the polarization used in the wave-particle kernel can be written as
\begin{equation*}
    \mathbf{E}=E_1\mathbf{e}_1+E_2\mathbf{e}_2+E_3\mathbf{b},
    \qquad
    E_1=1,\qquad
    E_2=\frac{i g}{\varepsilon-N^2},\qquad
    E_3=-\frac{N_\parallel N_\perp}{\eta-N_\perp^2}.
\end{equation*}
With this polarization convention, the cold-plasma wave-particle coupling factor is
\begin{equation*}
    U_l(\mathbf{p},\mathbf{k})
    =
    8\pi^2e^2
    \frac{
    \left[
    \frac{l\omega_c}{k_\perp v\gamma}J_l(\lambda)
    + E_3\cos\theta\,J_l(\lambda)
    + iE_2\sin\theta\,J_l'(\lambda)
    \right]^2
    }{
    \left(1-E_2^2\right)\frac{1}{\omega}\frac{\partial}{\partial\omega}\left(\omega^2\varepsilon\right)
    +2iE_2\frac{1}{\omega}\frac{\partial}{\partial\omega}\left(\omega^2g\right)
    +E_3^2\frac{1}{\omega}\frac{\partial}{\partial\omega}\left(\omega^2\eta\right)
    },
    \qquad
    \lambda=\frac{k_\perp v\gamma\sin\theta}{\omega_c}.
\end{equation*}
Here $J_l$ is the Bessel function of the first kind, $J_l'$ is its derivative with respect to $\lambda$, $v=p/(\gamma m)$, and $\cos\theta=p_\parallel/p$.

\section{Numerical methods}\label{sec:appendix_numerical_method}

The numerical implementation is a fixed-point iteration for the marginally stable runaway state described in the main text. We summarize here the implementation choices that are not fixed by the continuous model alone.

\paragraph{Leaking boundary at the origin.}
In practice we consider a truncated domain $\mathbf{p} \in \Omega \subset \mathbb{R}^{3}$. Suppose that $\overline{a}$ and $\overline{\overline{d}}$ are the advection vector and the diffusion tensor, we apply zero-flux condition on inflow regions and free flow condition on outflow regions:
\begin{equation*}
    \left(\overline{a}f -\overline{\overline{d}}\cdot\nabla_{p} f \right)\cdot \mathbf{n}^{-} = \begin{cases}
        0, &\overline{a}\cdot\mathbf{n}^{-} < 0\\
        \overline{a}f\cdot\mathbf{n}^{-}, & \text{otherwise.}
    \end{cases}
\end{equation*}

The origin requires a separate treatment because the normalized Coulomb drag is singular as $p\to0$:
\[
    \mathbf{F}_{c}=-\frac{\gamma^2}{p^2}\hat{\mathbf{p}}.
\]
Instead of imposing a reflecting condition at $p=0$, the solver lets particles that are dragged back to the thermal bulk leave the runaway calculation. A virtual inner boundary is introduced at small momentum $|\mathbf{p}|=\delta$, and the weak form becomes
\begin{equation*}
    \left(\frac{1}{p^{2}}\frac{\partial}{\partial p}\left(p^{2}\frac{\gamma^{2}}{p^{2}}f\right),\varphi\right)_{\mathbb{R}^{3} \setminus B(0,\delta)} =2\pi\int\left.\left(\gamma^{2}f\varphi\right)\right\vert_{|\mathbf{p}|=\delta}\sin\theta d\theta-\left(\frac{\gamma^{2}}{p^{2}}f,\frac{\partial\varphi}{\partial p}\right)_{\mathbb{R}^{3} \setminus B(0,\delta)}.
\end{equation*}
Let $\delta$ go to zero, we have 
\begin{equation*}
    \left(\frac{1}{p^{2}}\frac{\partial}{\partial p}\left(p^{2}\frac{\gamma^{2}}{p^{2}}f\right),\varphi\right)_{\mathbb{R}^{3} } =2\pi\int\left.\left(f\varphi\right)\right\vert_{|\mathbf{p}|=0}\sin\theta d\theta-\left(\frac{\gamma^{2}}{p^{2}}f,\frac{\partial\varphi}{\partial p}\right)_{\mathbb{R}^{3}},
\end{equation*}
and the first term on the right hand side characterizes the mass leaking rate to the bulk if we let $\varphi = 1$.

\paragraph{Fixed-point iteration.}

Adopting the self-consistent quasilinear diffusion operator $\mathbb{D}[f]$ at marginal stability, it remains to solve the following equation:
\begin{equation*}
\mathbb{E}f+\mathbb{C}f+\mathbb{Z}f+\mathbb{R}f+\mathbb{D}[f]f+\mathbb{S}f = 0.
\end{equation*}
Due to nonlinearity, the solution can only be obtained via a fixed-point iteration. We choose the one based on an implicit-explicit Euler scheme:
\begin{equation*}
    \begin{split}
        &\frac{f^{(n+1)} - f^{(n)}}{\Delta t} = \mathbb{E}f^{(n+1)}+\mathbb{C}f^{(n+1)}+\mathbb{Z}f^{(n+1)}+\mathbb{R}f^{(n+1)}+\mathbb{D}[f^{(n)}]f^{(n+1)}+\mathbb{S}f^{(n)}\\
        \Rightarrow \qquad & f^{(n+1)} = \left(\frac{1}{\Delta t}\mathbb{I} - \mathbb{E}-\mathbb{C}-\mathbb{Z}-\mathbb{R}-\mathbb{D}[f^{(n)}]\right)^{-1}\left(\frac{1}{\Delta t}\mathbb{I} +\mathbb{S}\right)f^{(n)}
    \end{split}
\end{equation*}
The iteration should be stopped when $f^{(n+1)}$ is close to $f^{(n)}$, in practice the integral $\int_{\mathbb{R}^{3}} \left[f^{(n+1)} - f^{(n)} \right]d^{3}\mathbf{p}$ is chosen as the criterion.

\paragraph{Role of the thermal-noise seed.}
The source $\Sigma$ represents a small thermal fluctuation level that seeds the waves. Its detailed functional form is not expected to affect the marginal state computed here. Away from marginality, $R<1$ and a small $\sigma$ produces only a small wave level. Near marginality, the factor $(1-R)^{-1}$, or its sigmoid regularization in the code, selects the wave-number region where the beam drive balances collisional damping. Thus a smooth positive seed changes only the small background level and the sharpness with which the spectrum turns on; it does not determine the ridge location, which is set by $R\simeq1$, nor the resulting wave-regulated runaway distribution in the small-noise limit.

\bibliographystyle{plain}
\bibliography{main.bib}
\end{document}